\documentclass[aps,pra,letterpaper,showpacs,superscriptaddress,twocolumn]{revtex4-1}

\usepackage{bbm}
\usepackage{amssymb}
\usepackage{ifpdf}
\usepackage{color}
\usepackage{graphicx}
\usepackage{amsmath,times,mathptm,txfonts}
\usepackage{amsfonts}
\usepackage{hyperref}
\usepackage{dcolumn}
\usepackage{bm}
\usepackage{xcolor}

\newcommand{\1}{{\mathbbm{1}}}

\newcommand{\ket} [1] {\vert #1 \rangle}
\newcommand{\bra} [1] {\langle #1 \vert}
\newcommand{\tr} {\textrm{Tr}}

\newcommand{\eqn}[1]{Eq.\ \ref{#1}}
\newcommand{\fig}[1]{Fig.\ \ref{#1}}

\definecolor{red}{rgb}{0.7,0,0}
\definecolor{green}{rgb}{0.,0.35,0.}
\definecolor{blue}{rgb}{0.2,0.2,0.7} 
\definecolor{black}{rgb}{0.15,0.15,.15}

\makeatletter
\@ifundefined{textcolor}{}
{%
 \definecolor{BLACK}{gray}{0}
 \definecolor{WHITE}{gray}{1}
 \definecolor{RED}{rgb}{1,0,0}
 \definecolor{GREEN}{rgb}{0,.4,0}
 \definecolor{BLUE}{rgb}{0,0,1}
 \definecolor{CYAN}{cmyk}{1,0,0,0}
 \definecolor{MAGENTA}{cmyk}{0,1,0,0}
 \definecolor{YELLOW}{cmyk}{0,0,1,0}
 }

\makeatother

\begin{document}

\title{Loops and strings in a superconducting lattice gauge simulator}
\author{G.K. Brennen}
\affiliation{Centre for Engineered Quantum Systems, Macquarie University, Sydney, NSW 2109, Australia}
\author{G. Pupillo}
\affiliation{icFRC, IPCMS (UMR 7504) and ISIS (UMR 7006), Universite de Strasbourg and CNRS, Strasbourg, France}
\author{E. Rico}
\affiliation{Department of Physical Chemistry, University of the Basque Country UPV/EHU, Apartado 644, E-48080 Bilbao, Spain}
\affiliation{IKERBASQUE, Basque Foundation for Science, Maria Diaz de Haro 3, E-48013 Bilbao, Spain}
\author{T.M. Stace}
\affiliation{Center for Engineered Quantum Systems, School of Mathematics and Physics, The University of Queensland, St Lucia, Queensland 4072, Australia}
\author{D. Vodola}
\affiliation{icFRC, IPCMS (UMR 7504) and ISIS (UMR 7006), Universite de Strasbourg and CNRS, Strasbourg, France}

\begin{abstract}
We propose an architecture for an analog quantum simulator of electromagnetism in $2+1$ dimensions, based on an array of superconducting fluxonium devices. The encoding is in the integer (spin-1)  representation of the quantum link model formulation of compact $U(1)$ lattice gauge theory. We show how to engineer  Gauss' law via an ancilla mediated gadget construction, and how to tune between the strongly coupled and intermediately coupled regimes. The witnesses to the existence of the predicted confining phase of the model are provided by non-local order parameters from Wilson loops and disorder parameters from \mbox{'t Hooft} strings. We show how to construct such operators in this model and how to measure them non-destructively via dispersive coupling of the fluxonium islands to a microwave cavity mode. Numerical evidence is found for the existence of the confined phase in the ground state of the simulation Hamiltonian on a ladder geometry. 
\end{abstract}
\pacs{ 03.67.Lx, 37.10.Ty, 71.38.Ht}
\maketitle

Gauge theories play a fundamental role in modern physics, including quantum electrodynamics and quantum chromodynamics. The discretised version of gauge theory, lattice gauge theory (LGT), is key to understand physics ranging from quantum spin liquids to quark-gluon plasmas \cite{Wilson:1973jt,rothe2012lattice,wen2004quantum}.  A fundamental phenomenon in gauge theories is the notion of confinement which manifests in the absence of isolated, colour-charged particles in nature, i.e., the only ``physical'' states are those that transform ``trivially'' under a gauge transformation. Yet quantum phases of gauge field theories cannot be characterised by local order parameters. Instead, non-local order parameters such as Wilson loops \cite{Wilson:1973jt} and \mbox{'t Hooft} strings  \cite{t-Hooft:1978} have been introduced to indicate the presence or absence of a confined phase. 

Quantum link models (QLM) provide a formulation of LGT's, in which finite-dimensional sub-systems associated with edges of the lattice encode the gauge field \cite{Horn,Orland,Chandrasekharan}. Related $U(1)$ gauge models are important for understand various condensed matter systems, including quantum spin ice models or quantum dimer models, which may exhibit deconfined critical points at $T=0$ \cite{Rokhsar:1988}. In principle QLM break Lorentz invariance while relativistic $U(1)$ gauge theories in $2+1$ dimensions are always in a confinement phase at $T=0$ but may undergo a phase transition at $T_{\text{c}} >0$ to a deconfined phase \cite{Polyakov:1987}. In either case, confinement physics is a key to understanding the phenomenology.
 
Numerical simulation of LGTs can be computationally costly due to the size of the Hilbert space or the sign problem with quantum Monte Carlo techniques \cite{Lin:2015} (for recent proposals using tensor networks see \cite{Byrnes,Banuls1,Banuls2,Rico,Buyens1,Silvi,Tagliacozzo2,Pichler,Buyens2,Haegeman,Kuhn,Zohar6,Zohar7}).  An alternative approach is to build analog quantum simulators to  replicate the equilibrium and dynamical properties of a system of interest.  Indeed, this is one of the motivations for quantum technologies based on atomic \cite{Weimer,Kapit,Zohar1,Casanova,Zohar2,Banerjee,Zohar3,Hauke,Tagliacozzo,Zohar4,Stannigel,Glaetzle,Notarnicola,Bazavov,Bermudez,Wiese,Zohar5} and superconducting  platforms \cite{Marcos1,Marcos2,Garcia,Mezzacapo}. A way to measure space-time Wilson loops in atomic lattice gauge simulators (assuming localised excitations) was given in Ref. \cite{Zohar3} but a critical outstanding problem has been the reliable measurement of non-local, space-like Wilson loops and~\mbox{'t Hooft} strings.

\begin{figure}[!t]
\begin{center}
\includegraphics[scale=0.35]{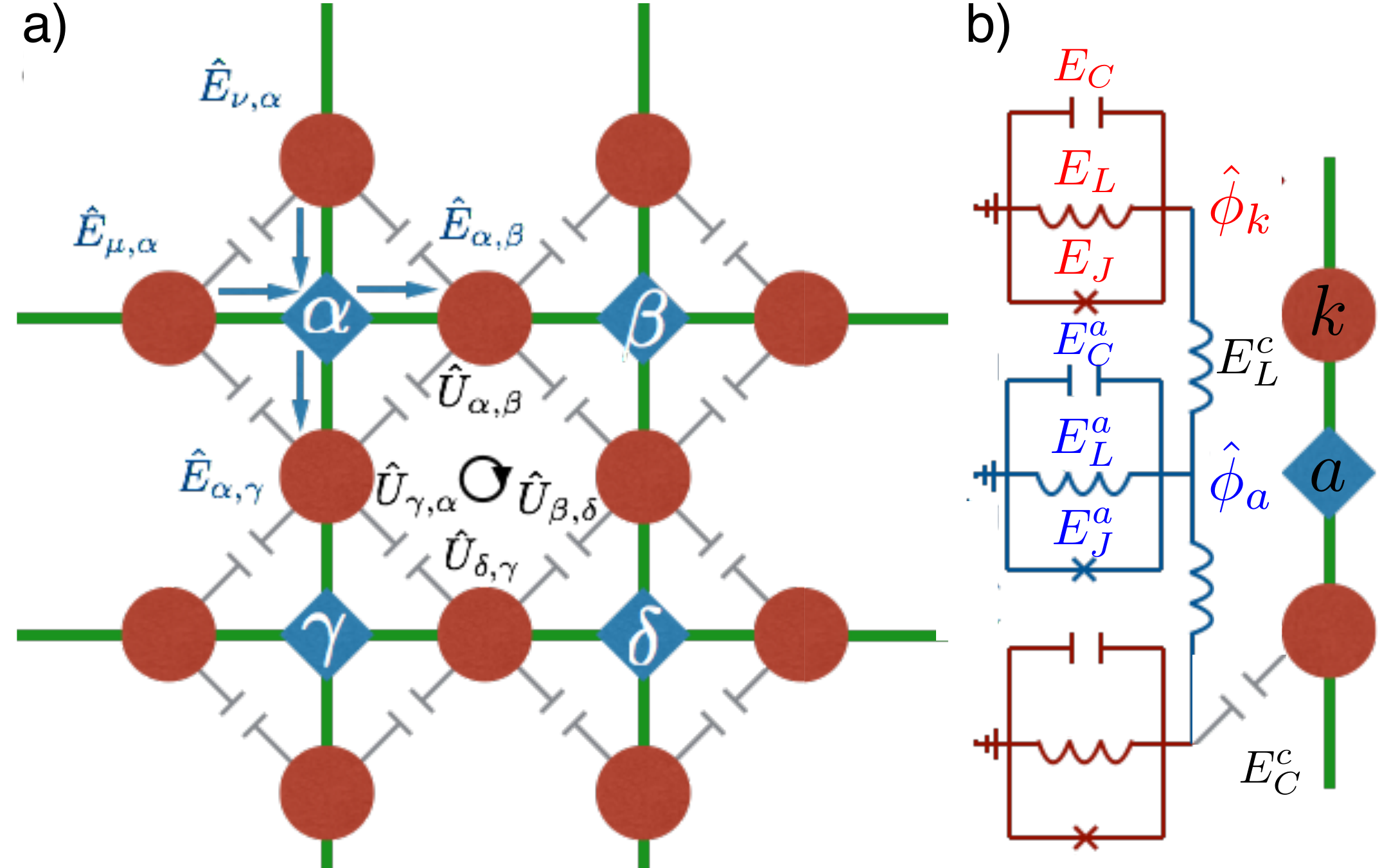}
\includegraphics[scale=0.37]{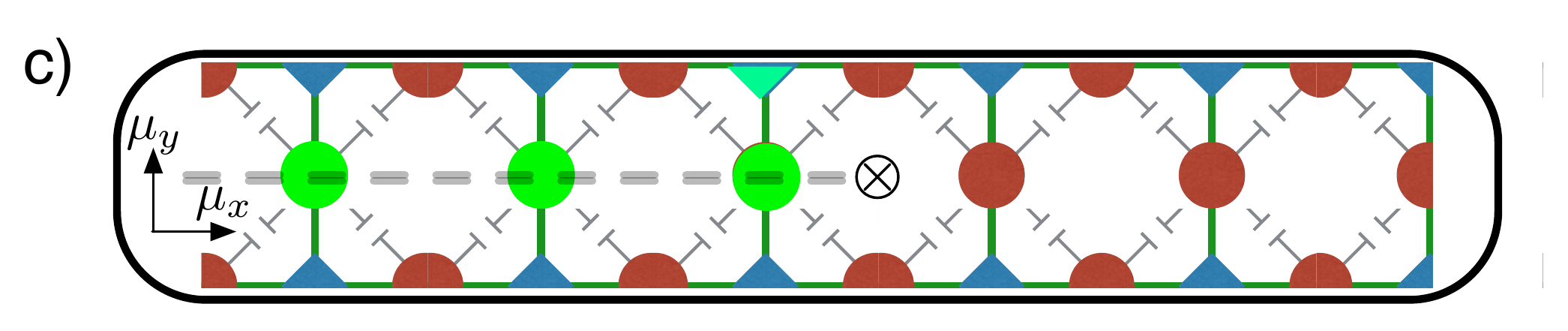}
\caption{$U(1)$ quantum link model engineered in a fluxonium array. {\bf a)} ``Electric", $\hat E_{\alpha,\beta}$, and ``magnetic", $\hat U_{\alpha,\beta}$, degrees of freedom are associated with links $\langle \alpha,\beta\rangle$ of a square lattice. Link degrees of freedom (red circles) are encoded in eigenstates of the fluxonia. Ancilla (blue diamonds) on vertices are inductively coupled to neighbouring link islands to mediate the Gauss constraint and plaquette interactions are obtained via link nearest neighbour capacitive coupling. {\bf b)} Superconducting circuit elements used to build and couple components of the simulation. The link devices have local phase $\hat{\phi}_{\rm link}$ and capacitive, inductive, and flux-biased Josephson energies $E_C$, $E_L$ and $E_J$ respectively and similarly for the ancilla devices. The capacitive and inductive coupling energies are $E^c_{C}$ and $E^c_L$. {\bf c)} A minimal quasi-1D `ladder' implementation  embedded in a microwave cavity (black box), in which a \mbox{'t Hooft} string of link fluxonia (green circles) can be measured via an ancilla coupled to the cavity (green triangle).}
\label{fig:piclgt1}
\end{center}
\end{figure}

Here, we propose an analog simulator of a pure compact $U(1)$ QLM in $2+1$ dimensions \cite{Kogut:1975zl}, based on superconducting fluxonium \cite{Manucharyan2009} devices placed  on a square lattice. The  devices operate in  a finite-dimensional manifold of low-lying eigenstates, to  represent `discrete' electric fluxes on the lattice. By engineering local couplings between  devices, we show how to replicate the local interactions and constraints of the QLM. The couplings can be tuned to access different phases of the quantum phase diagram of the model.  We demonstrate how to measure non-local, space-like Wilson loops and~\mbox{'t Hooft} strings in the proposed architecture. Moreover, we report density-matrix renormalisation group (DMRG) calculations of a \mbox{'t Hooft} disorder parameter in a QLM, and show that the QLM  indeed captures confinement physics.

\emph{Quantum Link Model:} In the pure gauge $U(1)$ QLM, electric fluxes, $\hat{E}_{\alpha, \beta}$, are defined on the links $\langle \alpha, \beta \rangle$ of a square lattice with local link state space $\mathbb{C}^{N+1}$, (circles in \fig{fig:piclgt1}a). In the `electric' basis,  the Hilbert space is labelled by the electric fluxes on the links, $\hat{E}_{\alpha, \beta} | E_{\alpha, \beta} \rangle = E_{\alpha, \beta} | E_{\alpha, \beta} \rangle$. For a compact $U(1)$ gauge group,  fluxes take integer or half integer values, \mbox{$-\frac{N}{2}\leq E_{\alpha, \beta}\leq \frac{N}{2}$}, $N\in\mathbb{Z^+}$. The local link electric-displacement operator, $\hat{U}_{\alpha, \beta}$, satisfies the commutation relation \mbox{$[ \hat{E}_{\alpha, \beta} , \hat{U}_{\alpha, \beta} ] = - \hat{U}_{\alpha, \beta}$} (for a detailed description see the Supplementary Information \cite{supp} (Sec.\ I)). In the charge-free sector, the net electric flux at a given vertex is zero, hence there is a conserved quantity $ \hat{G}_{\alpha}= \hat{E}_{\mu,\alpha} +  \hat{E}_{\nu,\alpha} - \hat{E}_{\alpha, \beta} -  \hat{E}_{\alpha,\gamma}$. The phase of the operators can be changed locally with the $U(1)$ gauge transformation $e^{ i  \theta_{\alpha}  \hat{G}_{\alpha} }$ and the dynamics remain invariant. The gauge invariant subspace satisfies $ \hat{G}_{\alpha} |\text{phys} \rangle =0$, which is  the discretised Gauss' law $\vec{\nabla} \cdot  \vec{E}\big|_{\text{phys}} = 0$. In a pure gauge model, there are two competing terms in the Hamiltonian: the electric term penalises electric flux on each link, $\langle \alpha,\beta\rangle$; and the magnetic term penalises magnetic flux on each plaquette $\square$,
\begin{equation}
\hat{H}_{\rm QLM} =  g^{2}_{\text{elec}}\hspace{-1mm} \sum_{\langle \alpha,\beta\rangle}    {\hat{E}_{\alpha,\beta}}^{2} - \frac{1}
{g^2_{\text{mag}}} \hspace{-0.5mm}\sum_{ \square}   (\hat{U}_{\alpha,\beta} \hat{U}_{\beta,\delta} \hat{U}_{\delta,\gamma} \hat{U}_{\gamma,\alpha}  + \textit{h.c.}) 
\label{puregaugeHam}
\end{equation}
where $g^{2}_{\text{elec}}$ and $g_{\text{mag}}^{2}$ are the coupling constants for the electric term and magnetic term, respectively.

We characterise the confinement of `electric' charges, which  locally violate Gauss' law, using Wilson loops. The smallest Wilson loop operator is defined on a plaquette \mbox{$\mathcal{W}=\hat{U}_{\alpha,\beta} \hat{U}_{\beta,\delta} \hat{U}_{\delta,\gamma} \hat{U}_{\gamma,\alpha}$}. This is a discrete approximation to $e^{i\sqint {\bf A}\cdot {\bf dl}}$ where ${\bf A}$ is the magnetic vector potential.   Over a longer closed path $\mathcal{C}$, a Wilson loop operator $\mathcal{W}_\mathcal{C}$ is the path-ordered multiplication of $\hat{U}_{\alpha,\beta}$ along links in $\mathcal{C}$. In the confined phase, $\mathcal{W}_\mathcal{C} $ satisfies an area law, $\langle \mathcal{W}_\mathcal{C} \rangle \sim e^{- \text{area}\left( \mathcal{C} \right)}$. In the deconfined phase, it satisfies $\langle \mathcal{W}_\mathcal{C} \rangle \sim e^{- \text{perimeter}\left( \mathcal{C} \right) }$. 

A \mbox{'t Hooft} string operator is defined as a directed product of electric link operators  $\hat{\Upsilon}(\varphi) = \prod_{n} \exp{( i \varphi \hat{E}_{n a_{x}, n a_{x} + a_{y}} )}$, in \cite{supp} (Sec.\ IB) we show that in the QLM it acts a \emph{disorder} parameter. This operator changes the value of the magnetic flux by an amount $\varphi$  on the plaquettes where it starts and ends, introducing a pair of magnetic vortices. In the confining phase it is ordered $\langle \hat{\Upsilon} (\varphi) \rangle \neq 0$ for $\varphi\neq 0$ in $2+1$ dimensions. The fact that a non-zero expectation value of the disorder parameter characterises a confinement phase in a gauge model may simplify the signal-to-noise problem in an actual quantum simulation.

\begin{figure}[!t]
\begin{center}
\includegraphics[scale=0.32]{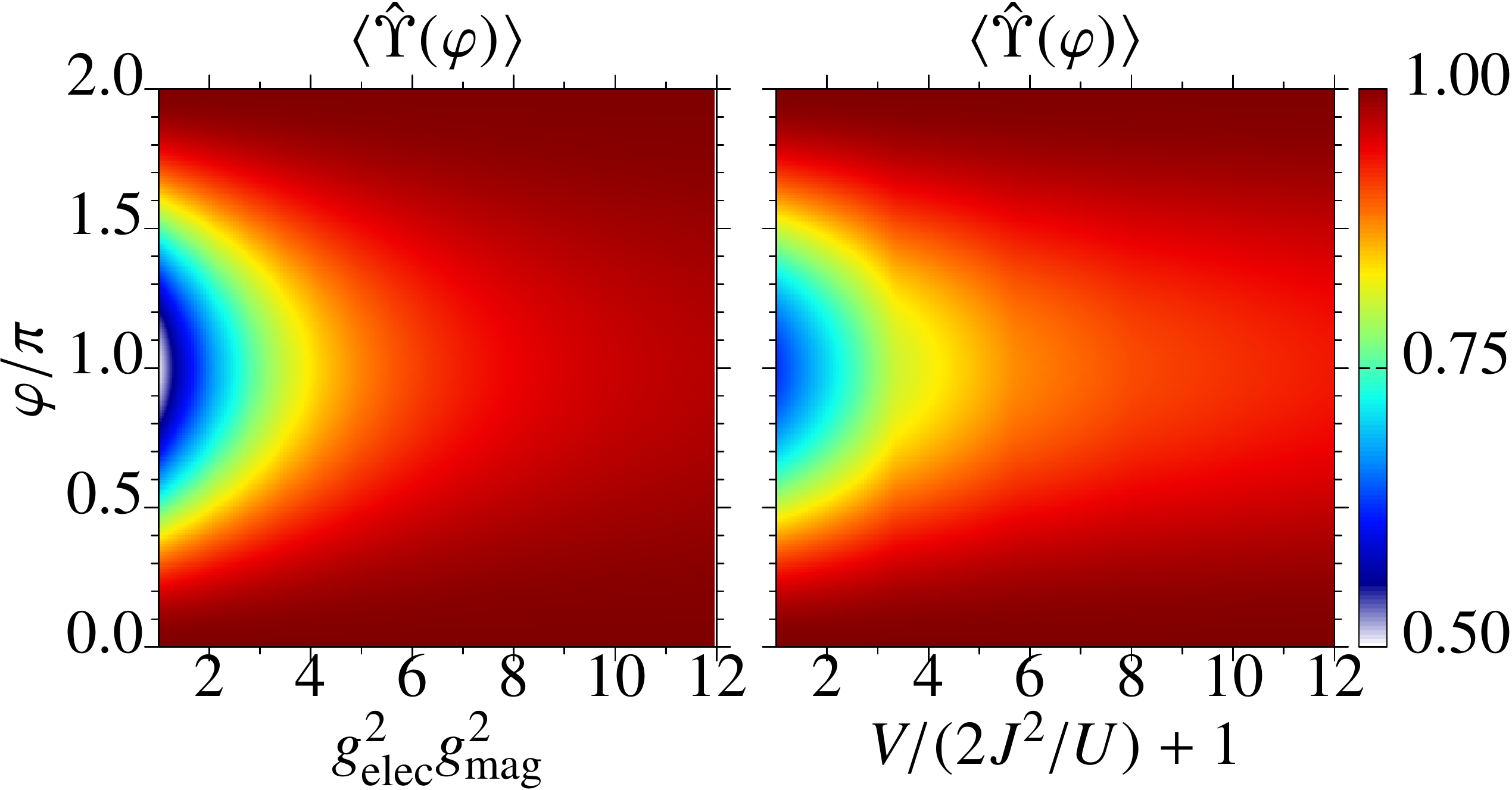}
\caption{Expectation value of the \mbox{'t Hooft} string $\hat{\Upsilon}(\varphi)$ which inserts a flux $\varphi$ at the plaquette in the middle of a ladder (see Fig. \ref{fig:piclgt1}c).  {\emph{(left panel)} Value in the ground state of the pure gauge model $\hat{H}_{\rm QLM}$ as a function of the electric coupling $g^{2}_{\text{elec}}$ with a perturbative value of the magnetic coupling $\frac{1}{g^2_{\text{mag}}} \sim \frac{2J^{2}}{U} \to \frac{2}{75}$. \emph{(right panel)}} Value in the ground state of the two-body Hamiltonian $\hat{H}_{\rm imp}$ as a function of the on-site energy $V$, with $J=1$ and $U=75$. Numerics were performed for a size $L=29$ rung ladder with $85$ spins  using DMRG with $300$ states and truncation error estimated at $<10^{-12}$. From the plots, $\langle \hat{\Upsilon}\rangle \ge 0.5$ is indicative of a confining phase. The equivalence between the implemented model, \mbox{\eqn{Hamimp}},  and the gauge invariant model, \mbox{\eqn{Hameff}}, in the strongly coupled and intermediately coupled regimes is evident.}
\label{fig:3}
\end{center}
\end{figure}

In \fig{fig:3}a we show the  disorder parameter, $\hat \Upsilon$,  for $\hat{H}_{\rm QLM}$ on the quasi-2D ladder lattice, shown in \fig{fig:piclgt1}c, calculated using DMRG. The ladder is the minimal lattice exemplifying a 2+1 dimensional system. Clearly, $\hat\Upsilon$ is non-zero in the strong coupling regime, $g^{2}_{\text{elec}} g^{2}_{\text{mag}}\gg1$, indicative of a confining phase. Thus, even in this limited geometry, the QLM  captures confinement physics.  In what follows, we propose an analog QLM simulator to study ground state and dynamical phenomena on computationally challenging 2D lattices.

\emph{Implementation with superconducting devices:} To  simulate a $U(1)$ QLM, there are three elements:
(i) The local Hilbert space,  labelled by the electric flux on the lattice links. Here, the Hilbert space is spanned by a discrete set of states of a \emph{fluxonium} device;
(ii)  Gauss' law on the lattice vertices.  Here, this is imposed by strong  interactions between devices, mediated by tunable inductive couplings;
(iii) The gauge invariant dynamics. Here, this emerges at second order of perturbation with capacitive couplings between neighbouring devices.

We propose a lattice of fluxonium devices \cite{Girvin:11}, which are inductively shunted superconducting Josephson junctions with demonstrated relaxation times on the order of $1ms$ \cite{Vool2014,Pop:2014}, located on the edges and vertices of the square lattice, as shown in \fig{fig:piclgt1}(a,b). The  Hamiltonian for device $k$ is  
\begin{equation}
\hat{H}_{k}=4 E_{C} \hat{n}_{k}^2+\hat{P}(\hat{\phi}_k),
\label{localHams}
\end{equation}
where \mbox{$\hat{P}(\hat{\phi}_k)= -E_{J}\cos(\hat{\phi}_{k}+\phi_{\rm off})+\displaystyle{E_{L}\hat{\phi}_{k}^2/{2}}$} is the local potential, $E_{C}={e^2}/{(2C)}$ is the charging energy of the island with total capacitance $C$, $E_{J}=(\frac{\hbar}{2e})^2\frac{1}{L_{J}}$ is the Josephson energy with $L_{J}$ the effective inductance of the Josephson junction,  $E_{L}=(\frac{\hbar}{2e})^2\frac{1}{L}$ is the shunt inductive energy, and $E_J \ge E_C > E_L$. 

\begin{figure}[!t]
\begin{center}
\includegraphics[width=0.9\columnwidth]{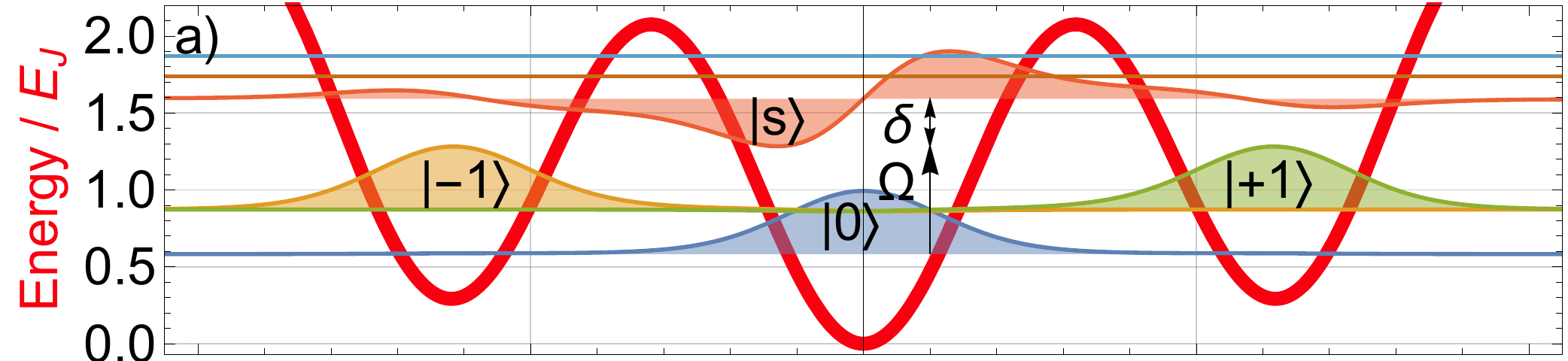}
\includegraphics[width=0.9\columnwidth]{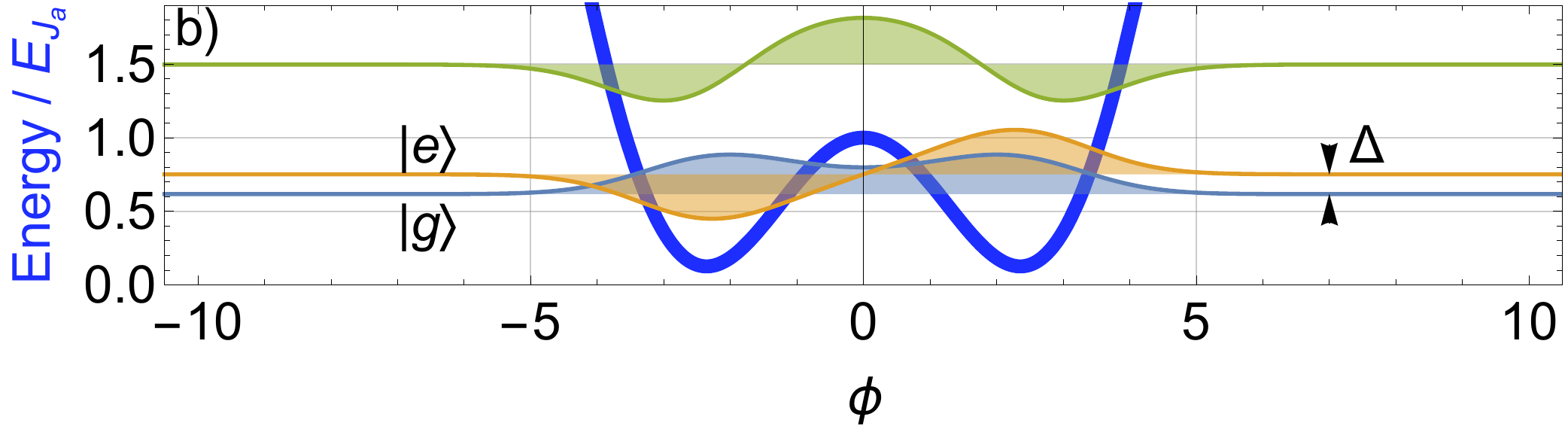}
\vspace{-5mm}
\caption{{\bf a)} Link fluxonium devices operate as qutrits with \mbox{$\phi_{\rm off}=0$}. The potential $P(\phi_{\rm link})$ (red) and the eigenfunctions are plotted for qutrit states $\ket{0},\ket{\pm1}$, which represent electric flux, and the next energy state $\ket{s}$. A cavity field couples states $\ket{0}\leftrightarrow \ket{s}$ with Rabi frequency $\Omega$ and detuning $\delta$ to tune the \emph{electric coupling} in the QLM. {\bf b)} Ancilla fluxonium potential $P(\phi_a)$ (blue) with $\phi_{\rm off}=\pi$, and eigenfunctions $\ket{g}$ and $\ket{e}$. Ancilla are initialised in state $\ket{e}$ and generate the \emph{Gauss constraint} on link devices through an effective interaction.}
\label{fig:qtrit}
\end{center}
\end{figure}

The  phase $\hat \phi_{k}$ is proportional to the (physical)  flux in the device.  It is not compact, so the conjugate charge $\hat{n}_{k}=-i\frac{\partial}{\partial \phi_{k}}$ takes continuous values. The offset phase $\phi_{\rm off}=2\pi{ \Phi_{\rm ext}}/{\Phi_{0}}$ where $\Phi_{\rm ext}$ is a tunable  flux \cite{Pop:2014} and $\Phi_0={h}/{ (2e)}$ is the  flux quantum. The potential terms can be tuned to support integer representations of the electric flux by setting $\phi_{\rm off}=0$, shown in \fig{fig:qtrit}a, or half-integer representations with $\phi_{\rm off}=\pi$, \mbox{\fig{fig:qtrit}b}.   In the limit $E_{J} \sim E_{C} \gg E_{L}$, the lowest energy states are the first band Wannier functions with mean (physical)  flux, \mbox{$\langle\hat \phi_{k}\rangle = 2\pi m_{k}$}, and zero-point phase fluctuations $\sigma_\phi = ( \frac{8E_{C}}{E_{J}})^{1/4}$. 

The \emph{Gauss law constraint} is enforced by ancillary fluxonium devices at  lattice vertices, with parameters $E^a_J,E^a_C,E^a_L$ and $\phi^a_{{\rm off}}=\pi$.  The lowest energy states $\ket{g}$ and $\ket{e}$ are shown in \fig{fig:qtrit}b. Each ancilla is inductively coupled to its neighbours, \mbox{$\hat{H}_{{\rm ind}} = (E^c_{L}/2) \sum_v\sum_{k=1}^4 (\hat{\phi}_k-\hat{\phi}_{a})^2$}, where \mbox{$E^c_{L}=(\frac{\hbar}{2e})^2\frac{1}{L_c}$}, and $L_c$ is the coupling inductance. Ancillae are initialised in the long-lived excited state $\ket{e}$, for which \mbox{$T_1>1$ms} \cite{Pop:2014}. 
 
 At first order in  $E^c_L/\Delta$, the interaction $\hat{H}_{\rm ind}$ is zero since $\ket{e}$ is antisymmetric. At second order we obtain an effective Hamiltonian acting on the links around each vertex. We only include the terms diagonal in the $\ket{m_k}$ basis since the off diagonal terms are much smaller by a factor $\sim e^{-\pi^2/\sigma^2}$. The total local plus inductive interaction for a spin$-S$ representation is then  $\sum_k \hat{H}_{k}+\hat{H}_{\rm ind}=V\sum_{k}{S^z_k}^2+U\sum_v(\sum_{k\in \mathcal{N}(v)}\hat{S}^z_k)^2$, where $\hat{S}^z=\sum_{m=-S}^S m\ket{m}\bra{m}$, \mbox{$U={E^c_L}^2 |\bra{g}\hat{\phi}_a\ket{e}|^2|\langle\hat{\phi}\rangle_{m=1}|^2/\Delta>0$}, $\mathcal{N}(v)$ is the neighbourhood of a vertex $v$  (see \cite{supp} Sec.\ IIB), \mbox{$\Delta=E_e-E_g>0$} is the ancilla qubit energy splitting including local contributions from the inductive coupling, calculated using \eqn{localHams}, with the replacement $E^a_L\rightarrow E^a_L+4E^c_L$, and $V$ (which generates the QLM \emph{electric coupling}, $g^2_{\rm elec}$) is the qutrit energy splitting $E_1-E_0$,  computed using \eqn{localHams} with  $E_L\rightarrow E_L+2E^c_L$. 

The QLM \emph{magnetic coupling}, $g^2_{\rm mag}$, is generated by a capacitive coupling between link devices, \mbox{$\hat{H}_{\rm cap} = 8E^c_C \sum_{\langle k,j \rangle} \hat{n}_k  \hat{n}_j$}, where $E^c_C/E_C\simeq \frac{\sqrt{8+\xi^{-2}}K_0(\xi^{-1})}{4\xi K(-16 \xi^2/(8+\xi^{-2}))}$, $\xi\equiv \sqrt{C_c/C}$ and $C_c$ is the capacitance between nearest neighbours, $K_0(x)$ is a modified Bessel function, and $K(x)$ is an elliptic integral (see \cite{supp} Sec.\ IIC). The operators $\hat{n}$ generate displacements in phase, so the interaction drives fluctuations in the electric flux $m_k$.  Longer range couplings decay exponentially in island separation, with a correlation length $\xi$. The total two body  Hamiltonian is 
\begin{equation}
\hat{H}_{\rm imp}=V\sum_{k}{S^z_k}^2+U\sum_v(\hspace{-1.5mm}\sum_{k\in \mathcal{N}(v)}\hspace{-1.5mm}\hat{S}^z_k)^2+J\hspace{-0.5mm}\sum_{\langle j,k\rangle}(\hat{S}_j^+ \hat{S}_k^- +\hat{S}_j^- \hat{S}_k^+),
\label{Hamimp}
\end{equation}
with $J=-8E^c_C|\bra{1}\hat{n}\ket{0}|^2$. In the limit $U \gg |J|$, the second term projects the ground states into the gauge invariant subspace and the effective Hamiltonian is  
\begin{eqnarray}
\hat{H}_{\rm eff}&=&(V + 2 {J^{2}}/{U} ) {\sum_{j}\hat{S}_j^z}^2-({2J^2}/{U})\sum_{\square}(\hat{S}^+ \hat{S}^-\hat{S}^+ \hat{S}^- +h.c.)\nonumber\\
&&{} + ({J^{2}}/{4 U}) \sum_{\langle j,k\rangle} \hat{S}_j^z \hat{S}_k^z (1 - \hat{S}_j^z \hat{S}_k^z ).\label{Hameff}
\end{eqnarray}
This is the first key element of our proposal: defining \mbox{$g^2_{\rm elec}=V+2J^2/U$} and $1/g^2_{\rm mag }=2J^2/U$, the first two terms of $\hat{H}_{\rm eff}$ are equivalent to the  gauge Hamiltonian $\hat{H}_{\rm QLM}$, \eqn{puregaugeHam}, once we identify $\hat{E} \mapsto \hat{S}^{z}$ and $\hat{U} \mapsto \hat{S}^{-}$, notice the later is non-unitary unlike the continuum case. The third term respects all the symmetries of the gauge invariant model, and renormalises the electric field and $U$.  Comparison of DMRG calculations of $\hat{H}_{\rm QLM}$ \mbox{(\fig{fig:3}a)} and $\hat{H}_{\rm eff}$ \mbox{(\fig{fig:3}b)} on a ladder geometry shows that $\hat{H}_{\rm eff}$ replicates the confinement physics ($\hat \Upsilon\neq0$) of the original QLM. We note  that higher order corrections to the gauge invariant Hamiltonian in Eq. \mbox{\ref{Hameff}} may lead to an effective coupling to dynamic matter field. In this case, the behaviour of the Wilson loops and \mbox{'t Hooft} strings in $2+1$D is uncertain, and left open to further study \cite{fradkinbook}.
 
We now discuss different limits of the model based on the three level (spin-1) $U(1)$ QLM  shown in \fig{fig:qtrit}a. The interaction energies are determined by  diagonalising the local fluxonium Hamiltonian and  computing wave-function overlaps (see \cite{supp} Sec.\ IID). We show how to prepare the strongly coupled limit, which is close to the product state $\otimes_{\rm links}\ket{0}_j$, and then reduce the coupling to the intermediate limit.

\emph{Strong coupling:} Because $\hat{H}_{\rm ind}$ contributes both to $U$ and $V$ in $\hat{H}_{\rm imp}$ with comparable magnitudes and $|J|\ll U$, in order to satisfy the Gauss constraint, the system will be in the strong coupling regime. To enforce this, the Josephson energy on the ancilla islands $E^{a}_{J}$ is the biggest energy scale of the model, which determines the cascade of energies: $E_{J} = E^{a}_{C} = 0.2 E^{a}_{J}$, $E_{C} = 0.06 E^{a}_{J}$, $E^{c}_{C}= 0.04 E^{a}_{J}$, $E^{a}_{L}=0.01 E^{a}_{J}$, $E_{L}=0.003E^{a}_{J}$ and $E^{c}_{L}=0.0002E^{a}_{J}$. The link states $\ket{\pm1}$ are then nearly degenerate with splitting $0.0006 E^{a}_J$ and $E^c_L/\Delta=0.023$ ensuring that the perturbation theory on the ancilla is valid. We find $U/E^{a}_J=0.006$, $V/E^{a}_J=0.06$, and $J/U=-0.04$, which by \eqn{puregaugeHam} gives $g^2_{\rm elec}g^2_{\rm magn}\sim 3000$.  Josephson energies $E_J=210$GHz \cite{Bylander:2011}, capacitive energies $E_C=14.2$GHz \cite{Vion:2002}, and inductive energies $E_L=0.52$ GHz \cite{MaslukThesis} have been reported, suggesting the simulation coupling strengths here are within reach of experimentally demonstrated values.

\emph{Intermediate  coupling:} To reduce the electric field term we  shift the energy of state $\ket{0}$ by off-resonant coupling to an excited state, $\ket{s}$, above the qutrit subspace. Consider a driving field that couples to the fluxonium at frequency $\omega_F$ which is detuned from the $\ket{0}\rightarrow \ket{s}$ transition by $\delta=\omega_F-(E_s-E_0)$. Because of the anharmonic energy spacing of the fluxonium the frequency, $\omega_F$ can be chosen to be very far off resonant for other possible transitions.  Inductive coupling via a quarter wavelength transmission line gives a time dependent fluxonium-field interaction is $\hat{H}_{FF}=\frac{\Omega}{2}\ket{s}\bra{0}e^{-i\omega_F t}+h.c.$ where the Rabi frequency is $\Omega=-g\bra{s}\hat{\phi}\ket{0}$ and $g$ depends on physical properties of the transmission line and fluxonium \cite{MaslukThesis}. Assuming other excited states are far-detuned,  $\hat{H}_{\rm imp}$ in the qutrit subspace is modified by $V\rightarrow V+{|\Omega|^2}/({4\delta})$, so $V$ can be reduced by choosing $\delta<0$. There are energy shifts  from off-resonant coupling to multiple excited states by all three qutrit states.  Optimising $\omega_F$ and  $g$ to minimise $V$ gives $\omega_F=1.588 E_J$ and $|g|^2=0.2$, so that $g^2_{\rm elec}g^2_{\rm mag}\simeq 1$.

\emph{Decoherence:} Spin decoherence limits the ultimate size of the simulator. We envision starting in the gapped product state $\ket{G(0)}=\prod_{\rm links}\ket{0}$ by tuning parameters to the extremely strongly coupled regime (with $\Omega=0$), and adiabatically evolving the ground state to an intermediate coupling regime. The adiabatic evolution could be done by slowly increasing the driving field Rabi frequency over a time $T_{\rm sim}$ and as described below, nonlocal order parameters can be measured as a function of final coupling strength (see \fig{fig:3}b). As shown in \cite{supp} (Sec.\ IIIB), the ground state $\ket{G(t)}$ is gapped throughout with energy $\Delta E_{\rm gap}(t)\sim 4g^2_{\rm elec}(t)$, and from the effective model $\hat{H}_{\rm eff}$ is minimal at $V=0$ where $\Delta E_{\rm gap}^{\rm min}\sim 8J^2/U$. The decoherence times for fluxonium tuned to the qutrit point has been reported at $T_1\sim T_2\sim 50 \mu$s \cite{Pop:2014}. Consider $U=0.032E_J$ as above and choose $E_J=40$GHz and $T_{\rm sim}=2/\Delta E_{\rm gap}^{\rm min}=0.135\mu$s. The inverted ancilla qubit lifetime is $T^a_1\sim 1$ms \cite{Pop:2014}, giving an error rate per ancilla of $1-e^{-T_{\rm sim}/T^a_{1}}\sim  10^{-4}$, allowing reliable simulations on a lattice with $\sim1000$ link spins. 

\emph{Nonlocal measurement:} The second key element of our proposal  is the  measurement of  spin-1 Wilson loop operators, \mbox{$\mathcal{W}_\mathcal{C}=\hat{S}^+\otimes \hat{S}^-\otimes \cdots \otimes \hat{S}^+\otimes \hat{S}^-$} on  $\mathcal{C}$, and \mbox{'t Hooft} disorder operators  \mbox{$\hat{\Upsilon}(\varphi)=e^{-i\varphi \hat{S}_0^z}\otimes e^{i\varphi \hat{S}_1^z}\otimes \cdots \otimes e^{i\varphi \hat{S}_{n-1}^z}$} on a line  extending from a spin $``0"$ on the boundary.  Importantly, the measurement does not alter the observable being measured, and repeated measurements give the same result, i.e.,\ it is \emph{non-demolition}. The idea is to prepare the ground state of the spin-1 lattice Hamiltonian, turn off $\hat{H}_{\rm imp}$, and then measure $\mathcal{W}_\mathcal{C}$ or $\hat{\Upsilon}(\varphi)$.  Thus, the measurement can `quench'  the system, in order to study the ensuing dynamics and multi-time correlations when the Hamiltonian is turned on again \cite{footnote}.  

To measure nonlocal operators, a subset of spins in the array are coupled to a single microwave cavity mode, \mbox{\fig{fig:piclgt1}c}. Ultimately, only a single qubit degree of freedom need be measured, which is advantageous if measurement error is significant. By contrast, if spins were measured independently the fidelity would decrease  exponentially with operator size. 

We require a dispersive coupling of spins in a region $\mathcal{R}$, $\hat{H}_{\rm int}=-\chi \hat{a}^{\dagger}\hat{a} \sum_{j\in\mathcal{R}}\ket{0}_j\bra{0}$, and a coupling between an ancilla $A$ (such as one of the ancilla qubits) and the bosonic field, $ \hat{H}^A_{\rm int}=-\chi_A \hat{a}^{\dagger}\hat{a}\ket{e}_A\bra{e}$, where $\hat{a}^{\dagger}$ and $\hat{a}$ are bosonic creation and annihilation operators. Selectively addressing cavity coupling within the region $\mathcal{R}$ or at the ancilla location can be done by coherently mapping  non-interacting spins to non-interacting local states which are far detuned from the cavity coupling. In \cite{supp} (Sec.\ III) we describe in detail two methods to measure  $\mathcal{W}_\mathcal{C}$ or $\hat{\Upsilon}(\varphi)$. In brief, one method uses a geometric phase gate, requiring only the ability to prepare the vacuum state of the cavity and a sequence of displacement operators and evolution generated by $\hat{H}_{\rm int}$. An alternative method can be done in a single step but requires the preparation of a superposition of vacuum and a single photon state of the cavity.

\emph{Fidelity:} To estimate process fidelity, we assume  the cavity has decay rate $\kappa$, and system and ancilla spins depolarise independently with error rate $\gamma$. On $n$ spins, the  geometric phase-gate measurement process-fidelity is
\begin{eqnarray}
F^{(gp)}_{\rm pro}(\theta,\Omega)&>&\eta_{A}(1-n(4\pi+6 \theta)\gamma/|\chi|) \label{gpfidelity}\\
&&\times\Big(1-{\pi\Omega\kappa} (e^{-3\theta\kappa/|\chi|}+e^{-\theta\kappa/|\chi|})\big(1+{\pi\kappa}/{2|\chi|}\big)/{|\chi|}\Big),\nonumber
\end{eqnarray}
where $\eta_{A}< 1$ describes the finite detection fidelity of the ancilla spin. For measuring Wilson loops, $\Omega=\pi/\sqrt{3}$, \mbox{$\theta=2\pi/3$}, while for measuring \mbox{\mbox{'t Hooft}} strings \mbox{$\Omega=\pi/2$}, \mbox{$\theta=\pi/2$}. For the single photon implementation, the process fidelity is \mbox{$F^{(sp)}_{\rm pro}(\theta)>\eta_{p}(1-n(1-e^{-\gamma\bar{t}(\theta)}))$}, where $\eta_{p}\leq 1$ describes finite single photon detection fidelity, and  the mean gate time is \mbox{$\bar{t}(\theta)=((1+e^{2\theta \kappa/|\chi|})(2\theta)^2\kappa/|\chi|^2)/(2\theta \kappa/|\chi|+e^{2\theta \kappa/|\chi|}-1)$}. In the presence of inhomogeneities in the dispersive coupling strength $\chi$, the error $\mathcal{E}$ for the global gates with angle $\theta$ is \mbox{$\mathcal{E}\approx\theta^2 |\mathcal{R}|(|\mathcal{R}|-1)\epsilon^2/2$} where $\epsilon$ is the  fractional cavity mode field variation across the lattice (\cite{supp} Sec.~IIIA). 

Using transmons coupled to a 3D microwave cavity \cite{Schuster:2011} the following values were reported for one island: $\gamma=66.7$kHz, $|\chi|/2\pi=99.8$MHz, $\kappa= 22.2$kHz. Single-shot transmon qubit measurements have also been reported with $\eta_A=0.919$  \cite{Houck}. With efficient single microwave photon detectors, the single photon protocol allows for a measurement of a Wilson loop on $n-1$ spins with fidelity $F_{\rm pro}^{(sp)}>\eta_p (1- 2.5\times 10^{-3}n)$. Microwave photon number resolution can be achieved with $\eta_p \simeq 0.90$ \cite{Girvin:10, Stace:14}. Assuming similar parameters for fluxonium and local addressability, using either the geometric phase gate or the single photon gate, a Wilson loop of length $8$ or a \mbox{'t Hooft} string of size $10$ could be measured with $\sim 90\%$ fidelity. By the non-destructive nature of the measurement, the imperfect detection efficiency can improved by repeating the measurement until the presence or absence of a photon is known with high confidence enabling measurement of much larger loops.

In summary, we provide a proposal for an analog 2+1D QLM simulator using a 2D array of superconducting devices. The simulator can be tuned between intermediate and strong coupling regimes, and allows non-destructive measurement of nonlocal,  space-like QLM order and disorder parameters, resolving an outstanding gap in other proposals. Moreover, we provide a physical encoding of the states for the QLM, where local electric field terms are non-trivial. The protocol is rather robust to inhomogeneities allowing for implementations in superconducting arrays, and we have presented numerical evidence that lattice QED in ``quasi-2''+1 dimensions exhibits confinement. Beyond ground state characterisation, the simulator can be used to probe dynamics and measure the evolution of non-local order/disorder parameters.

\emph{Acknowledgements}: This work was partially supported by the ARC Centre of Excellence for Engineered Quantum Systems EQUS (Grant No CE110001013). We also acknowledge financial  support from  Basque  Government  Grants  IT472-10, Spanish MINECO FIS2012-36673-C03-02,  UPV/EHU  Project  No. EHUA15/17,  UPV/EHU UFI 11/55 and SCALEQIT EU project. G.P. and D.V.  acknowledge  support  by  the  ERC-St  Grant  ColdSIM  (No.   307688),  EOARD,  UdS  via  Labex  NIE  and IdEX, RYSQ.

\begin{widetext}

\newpage

\pagebreak

\begin{center}
\textbf{\Large Supplemental Information}
\end{center}

In this supplemental material, we perform additional analysis that support the results obtained in the main article. 

\section{Large-$N$ representation of the $U(1)$ QLM and duality transformation}
\label{QLMreps}

The local Hilbert space in a $U(1)$ QLM is bounded and finite, i.e., $\mathcal{H}_{x,x+\mu_{x}} = \mathbb{C}^{N+1} = \{ | E_{x,x+\mu} \rangle  \}$, with $|E_{x,x+\mu}| \le \frac{N}{2}$ and $N \in \mathbb{Z}$. In this local Hilbert space, we redefine the local operators, such that
\begin{equation*}
\begin{split}
&\hat{E}_{x,x+\mu} | E_{x,x+\mu} \rangle =  E_{x,x+\mu} | E_{x,x+\mu} \rangle; \\
&\hat{W}_{x,x+\mu_{x}}=\frac{\hat{U}_{x,x+\mu_{x}}}{\sqrt{\frac{N}{2}  \left( \frac{N}{2} +1 \right) }}, ~ \, ~ \hat{W}_{x,x+\mu_{x}} | E_{x,x+\mu} \rangle = \sqrt{ \frac{\frac{N}{2} + 1 - E_{x,x+\mu}}{ \frac{N}{2} +1 } }  \sqrt{ \frac{ \frac{N}{2} + E_{x,x+\mu} }{ \frac{N}{2} } } | E_{x,x+\mu} -1\rangle;\\
&\hat{W}^{\dagger}_{x,x+\mu_{x}}=\frac{\hat{U}^{\dagger}_{x,x+\mu_{x}}}{\sqrt{\frac{N}{2}  \left( \frac{N}{2} +1 \right) }}, ~ \, ~\hat{W}^{\dagger}_{x,x+\mu_{x}} | E_{x,x+\mu} \rangle = \sqrt{ \frac{\frac{N}{2} + 1 + E_{x,x+\mu}}{ \frac{N}{2} +1 } }  \sqrt{ \frac{ \frac{N}{2} - E_{x,x+\mu} }{ \frac{N}{2} } } | E_{x,x+\mu} +1\rangle;
\end{split}
\end{equation*}
that fulfill the commutation relations
\begin{equation}
\begin{split}
& \left[ \hat{W}^{\dagger}_{x,x+\mu_{x}} , \hat{W}_{y,y+\mu_{y}}\right] =\delta_{x,y} \frac{2 \hat{E}_{x,x+\mu}}{ \frac{N}{2} \left( \frac{N}{2}  +1 \right)  }; \\
& \left[ \hat{E}_{x,x+\mu} , \hat{W}_{x,x+\mu_{x}} \right] = - \hat{W}_{x,x+\mu_{x}} 
\end{split}
\end{equation}

\begin{figure}[th]
\centering
\includegraphics[width=0.75\columnwidth]{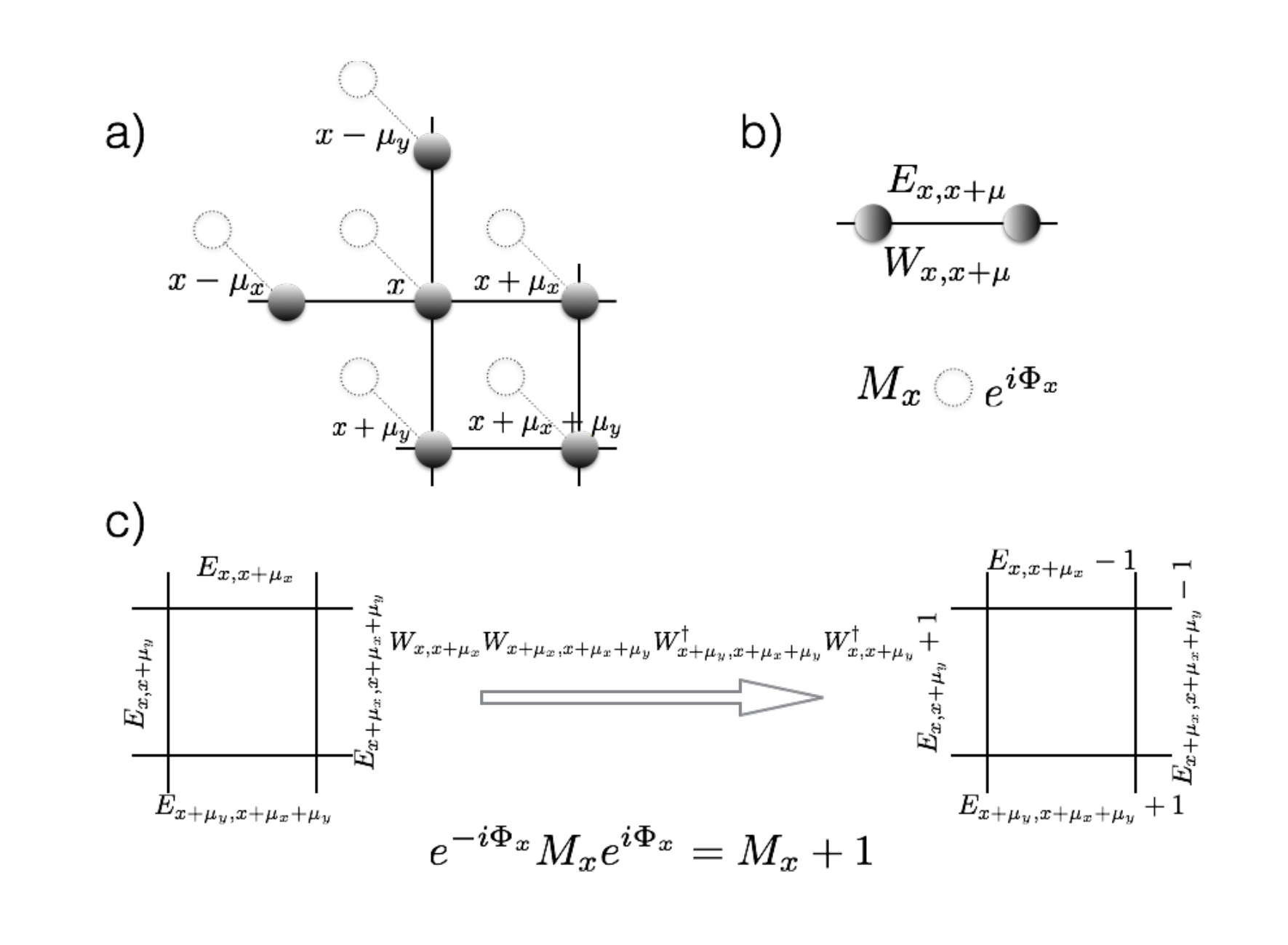}
\begin{caption}{\label{app1} a) Direct and dual lattice structure. b) Local degrees of freedom on the direct and dual models. c) Action of the plaquette operator on the direct and dual formulation.}
\end{caption}
\end{figure}

\subsection{Large-$N$ Hamiltonian} With the operators, we can define a large-$N$ Hamiltonian
\begin{equation}
\hat{\tilde{H}}  =  \hat{H}_{\text{elec}} + \hat{\tilde{H}}_{\text{magn}} =  g^{2}_{\text{elec}}  \sum_{x, \mu}    \left( \hat{E}_{x,x+\mu} \right)^{2} - \frac{1}{\tilde{g}_{\text{magn}}^{2}} \sum_{x, \mu_{x}, \mu_{y}} \left(  \hat{W}_{x,x+\mu_{x}} \hat{W}_{x+\mu_{x},x+\mu_{x}+\mu_{y}} \hat{W}_{x+\mu_{x}+\mu_{y},x+\mu_{y}} \hat{W}_{x+\mu_{y},x}  + \text{h.c.} \right),
\end{equation}
that is equivalent to Eq. (1) in the main text, if $g_{\text{magn}} = \tilde{g}_{\text{magn}} \left[ \frac{N}{2}  \left( \frac{N}{2} +1 \right) \right]^{2}$. In the limit $\langle \hat{E}_{x,x+\mu} \rangle \ll N$, the commutator of the links operators behaves as $  \left[ \hat{W}^{\dagger}_{x,x+\mu_{x}} , \hat{W}_{y,y+\mu_{y}}\right] \to 0 $ and we recover the Wilson limit.

In this representation, it is instructive to study the model defining the basic local degrees of freedom as ``plaquette'' variables, $e^{-i \hat{\Phi}_{x}} = \hat{W}_{x,x+\mu_{x}} \hat{W}_{x+\mu_{x},x+\mu_{x}+\mu_{y}} \hat{W}_{x+\mu_{x}+\mu_{y},x+\mu_{y}} \hat{W}_{x+\mu_{y},x}   $. Due to the previous commutation relations, in the large-$N$ limit, these local variables can be treated as independent. In addition to this, the operators given by 't Hooft strings, connecting a given plaquette with a reference one at infinity, correspond to local creation or annihilation operators on that plaquette. A complete description of the local ``dual'' Hilbert space and the dual Hamiltonian is the task of the next section.

\subsection{Dual Hamiltonian in the large-$N$ limit}
\label{DualHam}
 In what follows, we define the local dual Hilbert space and dual degrees of freedom, local dual operators, and the dual Hamiltonian dynamics of a $U(1)$ quantum link model in $2+1$ dimensions.

The local degrees of freedom are vortices located on every plaquette $x$ of the initial lattice. In the dual lattice the operator $e^{-i\hat{\Phi}_{x}} = \hat{W}_{x,x+\mu_{x}} \hat{W}_{x+\mu_{x},x+\mu_{x}+\mu_{y}} \hat{W}_{x+\mu_{x}+\mu_{y},x+\mu_{y}} \hat{W}_{x+\mu_{y},x}$, that measure the magnetic flux through a plaquette, is a local operator. 

Due to the Gauss' law around every vertex, there is a conservation of the total electric flux, i.e., $E_{x-\mu_{x},x} + E_{x-\mu_{y},x} = E_{x, x+\mu_{x}} + E_{x, x+\mu_{y}}$. It is fairly simple to realize that this relation is automatically satisfied if we define dual variables $M_{x}$ placed on the middle of the plaquette of the direct lattice (or on the sites of the dual lattice) such that:
\begin{equation}
\begin{split}
\hat{E}_{x,x+\mu_{y}} = \hat{M}_{x+\mu_{y}} - \hat{M}_{x+\mu_{x}+\mu_{y}}; ~ \, ~ \hat{E}_{x,x+\mu_{x}} = \hat{M}_{x+\mu_{x}+\mu_{y}} - \hat{M}_{x+\mu_{x}} ; ~ \, ~ \hat{E}_{x-\mu_{y},x} = \hat{M}_{x} -  \hat{M}_{x+\mu_{x}} ; ~ \, ~ \hat{E}_{x-\mu_{x},x} =   \hat{M}_{x+\mu_{y}} - \hat{M}_{x}.
\end{split}
\end{equation}
The solution of the dual variables in terms of the electric fluxes, reveals the variables $\hat{M}_{x}$ as the generators of the 't Hooft strings, i.e., $\hat{M}_{x} = \sum^{\infty}_{n=0} \hat{E}_{x-\mu_{y}+n\mu_{x},x+n\mu_{x}}$. In fact this dual variable and the magnetic flux are conjugate variables, fulfilling: $[\hat{M}_{x}, e^{\pm i \hat{\Phi}_{x}}] = \pm e^{\pm i \hat{\Phi}_{x}}$. 

With this change of variables, the dual Hamiltonian can be recast into
\begin{equation}
\hat{H}= g^{2}_{\text{elec}} \sum_{ x , \mu} \left[ \hat{M}_{x} - \hat{M}_{x+\mu} \right]^{2} - \frac{1}{\tilde{g}_{\text{magn}}^{2}} \sum_{x}  \left[ e^{i \hat{\Phi}_{x}} + e^{-i \hat{\Phi}_{x}}\right].
\end{equation}
This Hamiltonian is invariant under the unitary transformation $e^{i 2 \pi \sum_{x} \alpha_{x} \hat{M}_{x}}$ with $\alpha_{x} \in \mathbb{Z}$. Then, the spectrum of this Hamiltonian is periodic in $\Phi_{x}$ with a period $2\pi$ and the eigenfunctions $|\psi\rangle$ are Bloch's functions.

To analyze the behavior of this Hamiltonian, we study two limits $g^{2}_{\text{elec}} \tilde{g}_{\text{magn}}^{2} \gg 1$ and $g^{2}_{\text{elec}} \tilde{g}_{\text{magn}}^{2} \ll 1$. In the strong coupling limit $g^{2}_{\text{elec}} \tilde{g}_{\text{magn}}^{2} \gg 1$, the Hamiltonian reduces to $\hat{H} \to g^{2}_{\text{elec}} \sum_{ x , \mu} \left[ \hat{M}_{x} - \hat{M}_{x+\mu} \right]^{2} $, the eigenfunctions are eigenstates of the dual potential $\hat{M}_{x}$ that can be written as plane-waves in the magnetic flux basis, i.e., $\hat{M}_{x} | m_{x} \rangle = m_{x} | m_{x} \rangle$, $|m_{x} \rangle =\frac{1}{\sqrt{2\pi}} \int^{\pi}_{-\pi} d\phi_{x} ~ \, e^{i \phi_{x} m_{x}} |\phi_{x} \rangle$. In the weak coupling limit $g^{2}_{\text{elec}} \tilde{g}_{\text{magn}}^{2} \ll 1$, and the Hamiltonian reduces to $\hat{H} \to - \frac{1}{\tilde{g}_{\text{magn}}^{2}} \sum_{x}  \left[ e^{i \hat{\Phi}_{x}} + e^{-i \hat{\Phi}_{x}}\right]$.

From this duality exercise, we can realize that the previous gauge invariant $U(1)$ quantum link model is dual in the large-$N$ limit to a bosonic model with a global $U(1)$ invariance where the local degrees of freedom are magnetic vortices of the initial one. Hence, a possible phase transition in the gauge model is recast in a transition of the model with the global symmetry where the order parameter is given by the 't Hooft strings. The order phase corresponds to the confined phase, while the disorder phase in the global theory maps to a deconfined phase in the gauge theory.

\section{Details on the implementation with fluxonia}

\subsection{Local Hilbert space and the tight binding limit}

In a first case, we assume that the local inductive energy is negligible with respect to the local capacitive or the Josephson energies. The Hamiltonian for the fluxonium at link $k$ can be written $\hat{H}_k=\hat{H}^{J}_k+\frac{E_L}{2} \hat{\phi}_k^2$ where $\hat{H}^{J}_k=-4 E_{C}\frac{\partial^2}{\partial \phi_k^2}-E_J\cos(\hat{\phi}_k)$. In the tight binding limit, where the lattice wells are deep enough such that the particle dynamics is restricted to nearest neighbor only hoping, then one can find simplified expressions for the lowest energy eigenstates and eigenfunctions (see Ref. \cite{Pupillo:05}). For $E_L\ll E_J$ the lowest energy states $w_0(\phi_k-2\pi m_k)$ are the first band Wannier function with a well defined magnetic flux $2 \pi m_k$. For deep lattices we can approximate the Wannier functions as ground state harmonic oscillators whose wave-functions are approximately 
\[
w_0(\phi_k-2\pi m_k)=\frac{1}{(\pi \sigma^2)^{1/4}}e^{-(\phi_k-2\pi m_k)^{2}/2\sigma^2}.
\]
where 
\[
\sigma=\left(\frac{8E_{C}}{E_{J}}\right)^{1/4},
\]
is standard deviation of the wave-function localized at the potential minima $2\pi m$, $m_{\alpha,\beta}\in\mathbb{Z}$. The gap to the second energy band is of the order $\hbar\omega_{0}\sim \sqrt{8 E_{C} E_{J}} = \sigma^{2} E_{J}$ and the matrix elements of the Hamiltonian $\hat{H}^{J}_{\alpha,\beta}$ in this basis is:
\begin{equation}
\langle  m'_k | \hat{H}^{J}_k |m_k \rangle = E_{J} e^{-\pi^{2} \left(m'_k-m_k \right)^{2}/\sigma^{2}}  \left\{ \frac{\sigma^{2}}{4}  - \frac{\pi^{2} \left( m'_k- m_k\right)^{2}}{2} - e^{-\sigma^{2}/4} \cos{\left[ \pi (m'_k+m_k)\right]} \right\},
\end{equation}
and then the hoping rate $t$ between nearest neighbor wells is:
\begin{equation}
t= E_{J} e^{-\pi^{2} /\sigma^{2}} \left( \frac{\sigma^{2}}{4}  - \frac{\pi^{2} }{2} + e^{-\sigma^{2}/4} \right)
\end{equation}
The \emph{tight binding limit} is determined by the condition $\sigma^{2} E_{J} \gg | t | $  in which case we are justified in projecting onto the lowest energy level on the minimum of the cosine potential. This inequality is always fulfilled with the gaussian approximation. For instance, tunneling has a local extremum at  $\sigma \sim 3.168$ where $t \sim - 0.877 E_{J}$ and $\frac{\hbar\omega_{0}}{|t|} \sim 11.446$. In addition to this, the fact that we choose the minima at locations $\phi=2\pi m$ and all with the same standard deviation is valid only when the quadratic potential is a weak perturbation to the lattice, $E_L\ll E_{C} \le E_J$.

\subsection{Ancilla mediated inductive coupling for Gauss' Law constraint}
\label{ancillamediated}

Consider the interactions between link fluxonia which are inductively coupled pairwise, with energy $E^c_L$, to a single ancillary fluxonium $a$ located at a vertex $v$. The ancilla has an offset flux such that the lowest energy states define a qubit mode while for the links the low energy states define a qutrit. The Hamiltonian for this system is
\begin{equation}
\hat{H}_{\rm vertex} = \hat{H}_a +\sum_{k\in \mathcal{N}(v)}\hat{H}_k+ \frac{E^c_L}{2}\sum_{k\in \mathcal{N}(v)}(\hat{\phi}_a-\hat{\phi}_{k})^2 = \hat{H}_a' +\sum_{k\in \mathcal{N}(v)}\hat{H}_k'-E^c_L\hat{\phi}_a \sum_{k\in \mathcal{N}(v)}\hat{\phi}_{k},
\end{equation}
where $\mathcal{N}(v)$ denotes the set of links in neighborhood of the vertex $v$. Here $\hat{H}_a=E_e\ket{e}\bra{e}+E_g\ket{g}\bra{g}$ is the local ancilla Hamiltonian and $\hat{H}_k$ is the local link Hamiltonian without contributions from the inductive interaction.  The local Hamiltonians including the energy shifts due to the inductive couplings and assuming a square lattice with $|\mathcal{N}(v)|=4$ are 
\begin{equation*}
\begin{split}
\hat{H}'_a&=(E_e+2E^c_L\bra{e}\hat{\phi}_a^2\ket{e})\ket{e}\bra{e}+(E_g+2E^c_L\bra{g}\hat{\phi}_a^2\ket{g})\ket{g}\bra{g}, \\
\hat{H}'_k&=\hat{H}_k+\frac{E^c_L}{2} \sum_{\{m_k\}} \langle\hat{\phi}_k^2\rangle_{m_k}\ket{m_1m_2m_3m_4}\bra{m_1m_2m_3m_4}.
\end{split}
\end{equation*}
Define the ancilla energy splitting $\Delta=[(E_e+2E^c_L\bra{e}\hat{\phi}_a^2\ket{e})-(E_g+2E^c_L\bra{g}\hat{\phi}_a^2\ket{g})]>0$. Let's focus on the interaction part of the Hamiltonian. We transform to an interaction picture defined by $H_a'$, so that 
\begin{equation}
H_I(t)=-{E^c_L}\phi_a(t)\sum_{k\in \mathcal{N}(v)}\hat{\phi}_{k},
\end{equation}
where $\hat{\phi}_a(t)=e^{i \hat{H}_a' t}\hat{\phi}_a e^{-i \hat{H}_a' t}$.  To calculate the effective Hamiltonian induced by the ancilla, we need to expand the unitary evolution operator generated by $\hat{H}_I(t)$ to second order, project into the excited state $\ket{e}$ of the ancilla, and find the terms that increase linearly in $t$. The reason for projecting onto $\ket{e}$ for the ancilla rather than $\ket{g}$ is to arrive at the correct sign for the effective interaction as we show below. Specifically,
\begin{eqnarray}
\hat{U}_\textrm{eff}(t)&=&\tr_a \hat{U}(t) = \tr_a {\mathcal T}e^{-i \int_0^t dt \,\hat{H}_I(t)}\nonumber\\
&\approx&\bra{e}\left( 1-i \int_0^t dt_1 \,\hat{H}_I(t_1) - \int_0^t dt_1 \int_0^{t_1} dt_2 \,\hat{H}_I(t_1)\hat{H}_I(t_2)\right)\ket{e},\nonumber\\
&\equiv&1-i \,\hat{H}_\textrm{eff} t,\nonumber
\end{eqnarray}
where the last line defines $\hat{H}_\textrm{eff}$. Because the ancilla qubit states are eigenstates of the parity operator defined by $P\hat{\phi}=-\hat{\phi}$, we have $\bra{e}\hat{\phi}\ket{e}=0$, so the first order term is zero. Furthermore, as the inductive interaction is diagonal in the phase operator, coupling between different eigenstates $\ket{m}$ is exponentially suppressed. For example, in the tight binding limit, $|\bra{1}\hat{\phi}\ket{0}|/|\bra{1}\hat{\phi}\ket{1}|=e^{-\pi^2/\sigma^2}/2$. Hence the effective Hamiltonian is well approximated as diagonal in the $\ket{m}$ basis:
\begin{eqnarray}
-i\,\bra{m_1m_2m_3m_4}\hat{H}_\textrm{eff} \ket{m_1m_2m_3m_4} t&=&-\bra{e}\bra{m_1m_2m_3m_4} \int_0^t dt_1 \int_0^{t_1} dt_2 \,H_I(t_1)\ket{g}\ket{m_1m_2m_3m_4}\bra{g}\bra{m_1m_2m_3m_4}H_I(t_2)\ket{e}\ket{m_1m_2m_3m_4},\nonumber\\
&=&-\int_0^t dt_1 \int_0^{t_1} dt_2 \,\bra{e}\bra{m_1m_2m_3m_4}H_I(t_1)\ket{g}\ket{m_1m_2m_3m_4}\bra{g}\bra{m_1m_2m_3m_4}H_I(t_2)\ket{e}\ket{m_1m_2m_3m_4},\nonumber\\
&=&-{E^c_L}^2\int_0^t dt_1 \int_0^{t_1} dt_2 \, e^{i \Delta t_1}e^{-i \Delta t_2}|\bra{g}\phi_a\ket{e}|^2  |\bra{m_1m_2m_3m_4}\sum_{k\in \mathcal{N}(v)}\hat{\phi}_{k}\ket{m_1m_2m_3m_4})|^2  \nonumber\\
&\approx&-i \frac{{E^c_L}^2}{\Delta}  |\bra{g}\phi_a\ket{e}|^2 |\bra{m_1m_2m_3m_4}\sum_{k\in \mathcal{N}(v)}\hat{\phi}_{k}\ket{m_1m_2m_3m_4}|^2  t,\nonumber
\end{eqnarray}
where we assume $t\gg1/\Delta$, which is the appropriate regime for the effective Hamiltonian to make sense. This then gives us the following Hamiltonian around the vertex: 
\begin{equation}
\begin{array}{lll}
\hat{H}_{\rm vertex}
&=&\hat{H}'_a+\sum_{k\in \mathcal{N}(v)}\hat{H}_k'+\frac{ {E^c_L}^2 |\bra{g}\hat{\phi}_a\ket{e}|^2}{\Delta}\sum_{\{m_k\}} (\langle \hat{\phi}_1\rangle_{m_1}+\langle \hat{\phi}_2\rangle_{m_2}+\langle \hat{\phi}_3\rangle_{m_3}+\langle \hat{\phi}_4\rangle_{m_4})^2\ket{m_1m_2m_3m_4}\bra{m_1m_2m_3m_4}.
\end{array}
\end{equation}
In particular for a spin$-1$ encoding
\[
\hat{H}_{\rm vertex}=\hat{H}'_a+\sum_{k\in \mathcal{N}(v)}\hat{H}_k'+\frac{ {E^c_L}^2 |\bra{g}\hat{\phi}_a\ket{e}|^2|\langle\hat{\phi}\rangle_{m=1}|^2}{\Delta}(\sum_{k\in \mathcal{N}(v)}\hat{S}^z_k)^2,
\]
where we have used the fact that the expectation values satisfy $\langle\hat{\phi}\rangle_{m=1}=\langle\hat{\phi}\rangle_{m=-1}$ and $\langle\hat{\phi}\rangle_{m=0}=0$. On a square lattice, every link has contributions from interactions between two ancilla located at the vertices on the boundaries of the link. Hence the net effect of the modification of the local link terms is to make the replacement: $E_L\rightarrow E_L+2E^c_L$ and on the local ancilla: $E^a_L\rightarrow E^a_L+4E^c_L$.

Once we have the two previous ingredients: local Hilbert space spanned by well-defined magnetic fluxes and the local constraint that imposes zero flux on every vertex, any perturbative dynamics on this Hilbert space will be gauge invariant. We stress, once again, that the gap between the subspace with well-defined magnetic flux states and higher energy band is of order $\hbar\omega_{0}\sim \sqrt{8 E_{C} E_{J}} = \sigma^{2} E_{J}$. 

\subsection{Capacitive interaction}
\label{Capcalc}

Consider a square lattice with nodes at the midpoints of the edges, as in Fig.1a, in the main text. Suppose each node has capacitance to ground of $C$, and a capacitance $C_c$ to each of the neighboring nodes. The kinetic energy is
\begin{equation}
T=\sum_{\vec r} \left(\frac{C}{2} \, \dot\phi_{\vec r}^2+\sum_{{\vec j}_{\vec r}} \frac{C_c}{2} (\dot\phi_{\vec r}-\dot\phi_{{\vec j_{\vec r}}})^2\right),
\end{equation}
where the sum over ${\vec j}_{\vec r}$ is restricted to the 4 neighbors of node ${\vec r}$, and ${\vec r},{\vec j}_{\vec r}$ are vectors in the 2D lattice.

The conjugate charge is
\begin{equation}
q_{\vec r}=\partial_{\dot \phi_{\vec r}} T=(C+4C_c)\dot\phi_r-C_c\sum_{{\vec j}_{\vec r}}\dot\phi_{{\vec j}_{\vec r}},
\end{equation}
This implicitly defines the capacitance matrix $\mathbf{C}$ through \mbox{${\vec{q}}={\bf C}{\vec{\phi}}$} where ${\vec{q}}({\vec{\phi}})$ is the vector of charge(flux) degrees of freedom at the lattice points. Then we write the kinetic term as
\begin{eqnarray}
T=\frac{1}{2} \dot{\vec{\phi}}^T\mathbf{C}{\vec{\phi}} = \frac{1}{2} {\vec{q}}^T\mathbf{C}^{-1}{\vec{q}}.
\end{eqnarray}
Let's assume an $N_x\times N_y$ lattice with unit lattice spacing and periodic boundaries so that any lattice point can be indexed by an integer pair $\vec{r}=(r_x,r_y)\in\mathbb{Z}_{N_x}\times \mathbb{Z}_{N_y}$. The corrections to these results for lattices with open boundaries will scale like the inverse of the system size. The capacitance matrix can then be written
\[
\mathbf{C}=(C+4C_c){\bf 1}_{N_x N_y}-C_c ((X_{N_x}+X_{N_x}^{\dagger})\otimes {\bf 1}_{N_y}+{\bf 1}_{N_x}\otimes (X_{N_y}+X_{N_y}^{\dagger})),
\]
where the increment operator $X_N=\sum_{j=0}^{N-1}\ket{j+1}\bra{j}$. This can easily be diagonalized by a Fourier transform $F_N=\frac{1}{N}\sum_{j,k=0}^{N-1} e^{i2\pi jk/N}\ket{j}\bra{k}$ as
\[
F_{N_x}\otimes F_{N_y}\mathbf{C}F_{N_x}^{\dagger}\otimes F_{N_y}^{\dagger}=\sum_{n_x=0}^{N_x-1}\sum_{n_y=0}^{N_y-1} D({\vec k})\ket{n_x,n_y}\bra{n_x,n_y}.
\]
where $D({\vec k})=(C+4C_c)-2C_c(\cos(k_x)+\cos(k_y))$ and the momentum vectors are $\vec{k}=(k_x,k_y)=(\frac{2\pi n_x}{N_x},\frac{2\pi n_y}{N_y})$ where $(n_x,n_y)\in \mathbb{Z}_{N_x}\times \mathbb{Z}_{N_y}$. The inverse of the capacitance matrix is then
\[
\mathbf{C}^{-1}=F_{N_x}^{\dagger}\otimes F_{N_y}^{\dagger} \sum_{n_x=0}^{N_x-1} \sum_{n_y=0}^{N_y-1} \frac{\ket{n_x,n_y}\bra{n_x,n_y}}{D({\vec k})} F_{N_x}\otimes F_{N_y}.
\] 
and the kinetic energy is therefore,
\begin{equation}
T=\frac{1}{2}\sum_{{\vec r},{\vec j}} c_{{\vec r},{\vec j}}q_{\vec r}^* q_{{\vec j}},
\end{equation}
where
\[
\begin{array}{lll}
c_{{\vec r},{\vec j}}&=&\bra{r_x,r_y}F_{N_x}^{\dagger}\otimes F_{N_y}^{\dagger} \big[\sum_{n_x=0}^{N_x-1}\sum_{n_y=0}^{N_y-1}  \frac{\ket{n_x,n_y}\bra{n_x,n_y}}{D({\vec k})} \big]F_{N_x}\otimes F_{N_y}\ket{j_x,j_y}\\
&=&\frac{1}{N_xN_y}\sum_{n_x=0}^{N_x-1}\sum_{n_y=0}^{N_y-1}  \frac{e^{i {\vec k}\cdot ({\vec{r}-\vec{j})}}}{D({\vec k})}.
\end{array}
\]
In the large square lattice size limit $N_x\simeq N_y\rightarrow \infty$, for a separation between lattice sites $\Delta {\vec x}$
\begin{equation}
c({\Delta{\vec x}})=\frac{1}{(2\pi)^2}\int_{0}^{2\pi} dk_x \int_0^{2 \pi} dk_y \frac{e^{i\vec{k}\cdot \Delta{\vec x}}}{D(\vec k)}.
\end{equation}
Defining the quantity $\xi=\sqrt{C_c/C}$, the on-site interaction is 
\begin{equation}
c(0)=\frac{1}{C}\frac{2K\Big(-\frac{16\xi^2}{(\xi^{-2}+8)}\Big)}{\pi\xi \sqrt{\xi^{-2}+8}}\\
\end{equation}
where $K(x)$ is the elliptic integral of the first kind. In the limit $C_c\rightarrow 0$, $c(0)=1/C$ as expected. For $|\Delta{\vec x}|>0$ the non oscillatory support of the integrand is for $k\ll 1$ so we can expand the denominator as $D({\vec k})\approx C \xi^2 (k^2+\xi^{-2})$,
\begin{equation}
\begin{array}{lll}
c(|\Delta{\vec x}|>0)&\simeq&\frac{1}{C}\frac{1}{(2\pi\xi)^2}\int_{0}^{2\pi} dk k \int_0^{2 \pi} d\theta \frac{e^{i k|\Delta {\vec x}|\cos(\theta)}}{k^2+\xi^{-2}}\\
&=&\frac{1}{C}\frac{1}{2\pi\xi^2}\int_{0}^{2\pi} dk k \frac{J_0(k|\Delta{\vec x}|)}{k^2+\xi^{-2}}\\
&=&\frac{1}{C}\frac{1}{2\pi\xi^2}K_0(|\Delta{\vec x}|/\xi).
\end{array}
\end{equation}
where $J_0(x)$ is the zeroth Bessel function of the first kind and $K_0(x)$ is the modified zeroth Bessel function of the second kind. The parameter $\xi$ plays the role of a correlation length. For $|\Delta{\vec x}|/\xi\gg 1$, the interaction strength due to capacitive coupling between superconducting islands falls off exponentially with island separation $c(|\Delta{\vec x}|)\simeq\frac{1}{C} \sqrt{\frac{1}{8\pi |\Delta{\vec x}|/\xi}}\frac{e^{-|\Delta{\vec x}|/\xi}}{\xi^2}$. For short separations satisfying $|\Delta{\vec x}|/\xi\ll 1$, $c(|\Delta{\vec x}|)\simeq - \frac{\ln(|\Delta{\vec x}|/2\xi)+\gamma}{2\pi\xi^2}$, where $\gamma\approx 0.5772$ is the Euler gamma constant.

Under quantization of the charge degree of freedom $q_{\vec r}\rightarrow 2e\hat{n}_{\vec r}$ where $e$ is the electron charge and $\hat{n}_{\vec r}$ is the number operator for the Cooper pairs at the island at site ${\vec r}$, the kinetic energy operator is
\begin{equation}
T=4E_C\sum_{\vec r}\hat{n}_{\vec r}^2+ 8E^c_C \sum_{\langle {\vec j},{\vec r} \rangle} \hat{n}_{\vec j}  \hat{n}_{\vec r}+4E_C\sum_{{\vec j},{\vec r}; |{\vec j}-{\vec r}|>1}\frac{c(|{\vec j}-{\vec r}|)}{c(0)}\hat{n}_{\vec j}  \hat{n}_{\vec r},
\end{equation}
where the on site energy $E_C=\frac{e^2}{2C}$, and the nearest neighbor coupling magnitude is $E^c_C=E_C \frac{c(1)}{c(0)}\simeq E_C\frac{\sqrt{8+\xi^{-2}}K_0(\xi^{-1})}{4\xi K(-16 \xi^2/(8+\xi^{-2}))}$. The factor of two in the nearest neighbor term relative to the other terms is due to the fact that we are only count the interaction between the pair on a link once. Longer range interactions in the third term fall off exponentially with separation distance as described above. The matrix elements of the kinetic energy operator for between nearest neighbor islands are  computed by taking the wave-function overlaps of the charge operator in the local eigenbasis of fluxonium. In the tight binding limit we find:
\begin{equation}
\langle m'_{k} |\hat{n}_k | m_{k} \rangle = \frac{i}{\sigma^{2}} \pi \left( m'_{k} - m_{k} \right) e^{- \pi^{2}  \left( m'_{k} - m_{k}\right)^{2}/\sigma^{2}}.
\end{equation}

\subsection{Beyond tight binding limit: numerical calculation of local link eigenenergies and eigenfunctions}
\label{localeigen}
We wish to provide expressions for the interaction matrix elements that are valid even beyond the tight binding limit. This requires solving for the band structure for the Hamiltonian $\hat{H}_k$. First consider the local Hamiltonian with only the quadratic potential due to the inductance energy: $\hat{H}_k^{L}=-4 E_{C}\frac{\partial^2}{\partial \phi_k^2}+\frac{E_L}{2}\hat{\phi}^2_k$. The eigenstates and eigenfunctions are then simply those for a harmonic oscillator with origin of potential at $\phi_k=0$. Specifically, the eigenstates are labelled by $n\in\mathbb{N}$ with energies $E_n=\hbar\omega (n+\frac{1}{2})$, where $\hbar\omega=\sqrt{8 E_L E_C}$, and the eigenfunctions are 
\[
\psi_n(\phi_k)=\frac{1}{\sqrt{2^n n! \sqrt{\pi} \beta}} e^{-\phi_k^2/(2\beta^2)}H_{n}(\phi_k/\beta),
\]
where $\beta=(8 E_C/E_L)^{1/4}$, and $H_n(x)$ is the $n$-th Hermite polynomial. The Josephson term in the potential can then be computed in this basis. When the offset flux is set to $\phi_{\rm off}=0$ the matrix elements are
\[
\begin{array}{lll}
&&-E_J \bra{\psi_m(\phi_k)} \cos(\hat{\phi}_k) \ket{\psi_n(\phi_k)}=-E_J\frac{1}{\sqrt{2^{n+m} n! m!}} 2^{\min_{n,m}}(\min_{n,m})!(-1)^{|m-n|/2}\beta^{|m-n|}e^{-\beta^2/4}L_{\min_{n,m}}^{(|m-n|)}(\beta^2/2) \delta_{(m+n)\bmod 2,0},
\end{array}
\]
and when the offset flux is $\phi_{\rm off}=\pi$ the matrix elements are
\[
\begin{array}{lll}
&&-E_J \bra{\psi_m(\phi_k)} \sin(\hat{\phi}_k) \ket{\psi_n(\phi_k)}=-E_J\frac{1}{\sqrt{2^{n+m} n! m!}} 2^{\min_{n,m}}(\min_{n,m})!(-1)^{(|m-n|-1)/2}\beta^{|m-n|}e^{-\beta^2/4}L_{\min_{n,m}}^{(|m-n|)}(\beta^2/2) \delta_{(m+n)\bmod 2,1},
\end{array}
\] 
where $L_a^{(b)}(x)$ is the generalized Laguerre polynomial. Diagonalizing $\hat{H}_k$ in the basis $\{\psi_n(\phi_k)\}_{n=0}^{D-1}$ for some fixed dimension $D$ then gives approximations to the eigenenergies and eigenfunctions. In practice, for the regimes of interest, choosing $D\sim 80$ yields estimates of eigenenergies good to $\sim10^{-6}$ in units of $E_J$. 

\section{Measurement of Non-local order parameters}
\label{Nonlocalop}

\subsection{Measurement of Wilson loops}
\label{Wilson}
Consider the measurement of the Wilson loop operator for the $U(1)$ QLM using spin$-1$ particles.  We want a way to measure loop operators like
\[
\hat{W}(\mathcal{C})=\hat{S}^+\otimes \hat{S}^-\otimes \cdots \otimes \hat{S}^+\otimes \hat{S}^-,
\]
on a spatial loop $\mathcal{C}$ on the lattice where
\begin{equation}
S^+=\sqrt{2}\left(\begin{array}{ccc}0 & 1 & 0 \\0 & 0 & 1 \\0 & 0 & 0\end{array}\right)=(S^-)^{\dagger}.\label{eqn:splus}
\end{equation}
We will make use of the unitary Pauli $\hat{X}$ operator on qutrits:
\[
\hat{X}=\left(\begin{array}{ccc}0 & 1 & 0 \\0 & 0 & 1 \\1 & 0 & 0\end{array}\right)=\frac{\hat{S}^+}{\sqrt{2}}+\frac{(\hat{S}^-)^2}{2}.
\]
The weighted expectation value of the unitary loop operator of length $|\mathcal{C}|$ (which is even on a square lattice) is
\[
\begin{array}{lll}
2^{|\mathcal{C}|/2}\langle \hat{X}\otimes \hat{X}^{\dagger}\otimes \cdots \otimes \hat{X}\otimes  \hat{X}^{\dagger}\rangle = \langle (\hat{S}^+ +\frac{(\hat{S}^-)^2}{\sqrt{2}})\otimes (\hat{S}^- +\frac{(\hat{S}^+)^2}{\sqrt{2}})\otimes \cdots \otimes (\hat{S}^+ +\frac{(\hat{S}^-)^2}{\sqrt{2}})\otimes (\hat{S}^- +\frac{(\hat{S}^+)^2}{\sqrt{2}})  \rangle.
\end{array}
\]
If we restrict to the gauge invariant space of states then the expectation value takes the simple form
\[
\begin{array}{lll}
\langle \hat{V}(\mathcal{C})\rangle \equiv \langle \hat{X}\otimes \hat{X}^{\dagger}\otimes \cdots \otimes \hat{X}\otimes  \hat{X}^{\dagger}\rangle =2^{-|\mathcal{C}|/2}\langle \hat{W}(\mathcal{C})\rangle+2^{-|\mathcal{C}|}\langle(\hat{W}^{\dagger}(\mathcal{C}))^2\rangle.
\end{array}
\]
We can cancel the additional term using the modified unitary operator 
\[
\hat{X}(\phi)=\left(\begin{array}{ccc}0 & 1 & 0 \\0 & 0 & 1 \\e^{i\phi} & 0 & 0\end{array}\right)=\frac{\hat{S}^+}{\sqrt{2}}+e^{i\phi}\frac{(\hat{S}^-)^2}{2},
\]
on one lattice spin (say the first spin) such that In the gauge invariant sector
\[
\begin{array}{lll}
\langle \hat{V'}(\mathcal{C})\rangle \equiv \langle \hat{X}(\pi)\otimes \hat{X}^{\dagger}(0)\otimes \cdots \otimes \hat{X}(0)\otimes  \hat{X}^{\dagger}(0)\rangle = 2^{-|\mathcal{C}|/2}\langle \hat{W}(\mathcal{C})\rangle-2^{-|\mathcal{C}|}\langle (\hat{W}^{\dagger}(\mathcal{C}))^2\rangle.
\end{array}
\]
The sum of expectation values is $\langle \hat{V}(\mathcal{C})\rangle+\langle \hat{V'}(\mathcal{C})\rangle=2^{-|C|/2}\langle \hat{W}(\mathcal{C})\rangle$.

Measurement of $\hat{V}(\mathcal{C})$ can be done by coupling the lattice of spin to a bosonic mode \cite{Jiang:07}. We first consider a realization of the measurement via a geometric phase gate using coherent state displacements of the mode, and second a realization using a single Fock state excitation.    

In order to measure $\hat{V}(\mathcal{C})$ it is convenient to transform to a basis diagonal in the spin states:
\[
\hat{V}(\mathcal{C})=\hat{X}\otimes \hat{X}^{\dagger}\otimes \cdots \otimes \hat{X}\otimes  \hat{X}^{\dagger}=\hat{W}^{\dagger} \hat{Z}^{\otimes |\mathcal{C}|} \hat{W},
 \]
 where 
 \[
\hat{Z}=\left(\begin{array}{ccc}1 & 0 & 0 \\0 & \xi & 0 \\0 & 0 & \xi^2\end{array}\right),
 \]
 $\hat{W}=\hat{F}^{\dagger}\otimes \hat{F} \otimes \cdots \otimes \hat{F}^{\dagger}\otimes \hat{F}$, and the discrete Fourier transform is $\hat{F}=\frac{1}{\sqrt{3}}\sum_{r,s=0}^2\xi^{rs} \ket{r}\bra{s}$ with $\xi=e^{i2\pi/3}$.  Hence we can focus on generating evolution by the diagonal many body operator $\hat{Z}^{\otimes |\mathcal{C}|}$.
 
\subsubsection{Geometric phase gate}
\label{Geophase}
Let $\hat{a}$ and $\hat{a}^{\dagger}$ be creation and annihilation operators for a bosonic mode satisfying $[\hat{a},\hat{a}^{\dagger}]=1$.  The displacement operator acting on this mode is $\hat{D}(\alpha)=e^{\alpha \hat{a}^{\dagger}-\alpha^* \hat{a}}$, and the phase rotation operator is $\hat{R}(\theta)=e^{i\theta \hat{a}^{\dagger}\hat{a}}$. For any operator $\hat{O}$ acting on the lattice spins note the following identity
\begin{equation}
\hat{D}(\alpha e^{i\phi+i\theta \hat{O}})=\hat{R}(\theta \hat{O})\hat{D}(\alpha e^{i\phi})\hat{R}(-\theta \hat{O}).
\label{displacement}
\end{equation}
The following sequence produces a geometric phase gate on the spins which corresponds to evolution by an effective many-body Hamiltonian \cite{Jiang:07,Brennen:09}
\begin{equation}
\begin{array}{lll}
\hat{U}(\phi,\theta,\Omega)=\hat{D}(-\beta)\hat{R}(\theta \hat{O})\hat{D}(-\alpha)\hat{R}(-\theta \hat{O})\hat{D}(\beta) \hat{R}(\theta \hat{O})\hat{D}(\alpha)\hat{R}(-\theta \hat{O}) =\exp(-i\Omega\sin(\theta\hat{O}+\phi)),
\label{unitary}
\end{array}
\end{equation}
where $\phi=\arg(\alpha)-\arg(\beta)$ and $\Omega=|\alpha\beta|$.  Note after this sequence, the bosonic mode is disentangled from the spins.  In particular, if it begins in the vacuum then it ends in the vacuum state, and the first controlled phase rotation $\hat{R}(-\theta \hat{O})$ in Eq. (\ref{unitary}) can be omitted. Choose $\hat{O}$ to be a sum over local operators on the spins along the loop $\mathcal{C}$: $\hat{O}=\sum_{j\in\mathcal{C}}\sum_{k=1}^2 k\ket{k}_j\bra{k}$, where $\ket{0}=\ket{\hat{S}^z=1},\ket{1}=\ket{\hat{S}^z=0},\ket{2}=\ket{\hat{S}^z=-1}$. The physical interaction required to generate the unitary $\hat{R}(\theta\hat{O})$ is dispersive coupling between the bosonic mode and the spins
 \begin{equation}
\hat{H}_{\rm int}=-\chi \hat{a}^{\dagger}\hat{a} \sum_{j\in\mathcal{C}}\sum_{k=1}^2 k\ket{k}_j\bra{k}.
 \end{equation}
Specifically, $\hat{R}(\theta\hat{O})=e^{-i\hat{H}_{\rm int}\tau}$ for $\tau=\theta/\chi$.  It practice it may be simpler to implement in two steps with dispersive coupling to a single basis state $\ket{0}$ in each step: $\hat{R}(\theta\hat{O})=F_{1}^{\dagger}e^{i\chi\tau_1 \hat{a}^{\dagger}\hat{a} \sum_{j\in\mathcal{C}} \ket{0}_j\bra{0}}F_1F_2^{\dagger}e^{i\chi\tau_2 \hat{a}^{\dagger}\hat{a} \sum_{j\in\mathcal{C}} \ket{0}_j\bra{0}}F_2$ with $\tau_2=2\tau_1=\theta/\chi$ and the local state permutation unitaries $F_1=\ket{0}\bra{1}+\ket{1}\bra{0}+\ket{2}\bra{2}$ and $F_2=\ket{0}\bra{2}+\ket{2}\bra{0}+\ket{1}\bra{1}$.

There following parameter settings in the geometric phase gate is of interest:
\[
\hat{U}(0,\frac{2\pi}{3},\frac{\pi}{\sqrt{3}})=\frac{1}{3}{\bf 1}+(\frac{1}{3}-\frac{1}{\sqrt{3}})\hat{Z}^{\otimes |\mathcal{C}|}+(\frac{1}{3}+\frac{1}{\sqrt{3}})\hat{Z}^{\dagger\otimes |\mathcal{C}|}.
 \]

Now we also need a coupling between an ancilla spin $A$ and the bosonic field 
 \[
\hat{H}^A_{\rm int}=-\chi \hat{a}^{\dagger}\hat{a}\ket{1}_A\bra{1}.
 \]
This will enable the gate $\hat{R}(\pi \ket{1}_A\bra{1})=e^{-i\hat{H}^A_{\rm int} \tau}$ for $\tau=\pi/\chi$ which can be used to obtain a controlled displacement of the bosonic mode dependent on the state of the ancilla state:
 \[
 \begin{array}{lll}
\hat{D}(\beta \ket{1}_A\bra{1})= \hat{D}(\beta/2)\hat{R}(\pi \ket{1}_A\bra{1}) \hat{D}(-\beta/2)R(-\pi \ket{1}_A\bra{1}) =\ket{0}_A\bra{0}\otimes {\bf 1}+\ket{1}_A\bra{1}\otimes \hat{D}(\beta).
 \end{array}
 \]
The following composition of the these primitive gates will enable a controlled many-body unitary conditioned on the state of the ancilla:
 \begin{equation}
 \begin{array}{lll}
 \hat{M}(\phi,\theta,\Omega)=\hat{D}(-\beta \ket{1}_A\bra{1})\hat{R}(\theta \hat{O})\hat{D}(-\alpha)\hat{R}(-\theta \hat{O})\hat{D}(\beta \ket{1}_A\bra{1}) \hat{R}(\theta \hat{O})\hat{D}(\alpha)\hat{R}(-\theta \hat{O}) =\ket{0}_A\bra{0}\otimes {\bf 1}+\ket{1}_A\bra{1}\otimes \hat{U}(\phi,\theta,\Omega).
\label{controlledunitary}
\end{array}
 \end{equation}
Beginning with the ancilla in the state $\ket{+_x}_A=\frac{1}{\sqrt{2}}(\ket{0}+\ket{1})$ and the lattice spins in the state $\ket{\psi}$, perform local rotations using $W$ followed by the controlled many-body unitary, followed by the inverse local rotations:
 \[
\hat{W}^{\dagger} \hat{M}(\phi,\theta,\Omega)\hat{W}\ket{\psi}\ket{+_x}_A=\frac{1}{\sqrt{2}}(\ket{\psi}\ket{0}_A+\hat{W}\hat{U}(\phi,\theta,\Omega)\hat{W}^{\dagger}\ket{\psi}\ket{1}_A).
 \]
The measured polarization of the ancilla along the $\hat{x}$ and $\hat{y}$ directions is 
\[
\begin{array}{lll}
\langle \hat{\sigma}_A^x\rangle(\phi,\theta,\Omega)=\Re[\langle \hat{W}\hat{U}(\phi,\theta,\Omega)\hat{W}^{\dagger}\rangle]\\
\langle \hat{\sigma}_A^y\rangle(\phi,\theta,\Omega)=\Im[\langle \hat{W}\hat{U}(\phi,\theta,\Omega)\hat{W}^{\dagger}\rangle],\\
\end{array}
\]
where the expectation value on the right hand side is taken with respect to the spin lattice many-body state $\ket{\psi}$. For the aforementioned geometric phase gate parameters
\[
\begin{array}{lll}
\langle \hat{\sigma}_A^x\rangle(0,\frac{2\pi}{3},\frac{\pi}{\sqrt{3}})=\frac{1}{3}(1+2\times 
\Re[\langle \hat{V}(\mathcal{C})\rangle])\\
\langle \hat{\sigma}_A^y\rangle(0,\frac{2\pi}{3},\frac{\pi}{\sqrt{3}})=-\frac{2}{\sqrt{3}}
\Im[\langle \hat{V}(\mathcal{C})\rangle].
\end{array}
\]
Hence we can perform a non-demolition measurement of the real and imaginary parts of $V(\mathcal{C})$.   In order to measure $V'(\mathcal{C})$, we use a very similar protocol but with the replacements:  $\hat{W}\rightarrow \hat{W'}=\hat{F}^{'\dagger}\otimes \hat{F}\cdots\otimes \hat{F}^{'\dagger}\otimes \hat{F}$ where
\[
\hat{F'}=\frac{1}{\sqrt{3}}\left(\begin{array}{ccc}1 & \xi & \xi^2 \\ -1 & -\xi^2 & -\xi \\1 & 1 & 1\end{array}\right),
\]
acts on every other (say even labelled) spins, and choose the interaction between the spin and the bosonic mode to have the opposite sign, $\chi\rightarrow -\chi$, for the even labelled spins, which is equivalent to changing $\theta\rightarrow -\theta$ for the unitary evolution on those spins.  This enables a non-demolition measurement of the Wilson loop.

\subsubsection{Single photon mediated gate}
\label{singlephoton}
If single Fock states of the bosonic mode can be prepared then the measurement protocol is significantly simplified.  Consider a single Fock state of a photonic mode which is a superposition of polarization modes:  $\ket{\phi}_{\rm field}=\frac{1}{\sqrt{2}}(\hat{a}^{\dagger}_++\hat{a}^{\dagger}_-)\ket{vac}=\frac{1}{\sqrt{2}}(\ket{1}_++\ket{1}_-)$.  Also let the lattice spins interact with the photon via a polarization dependent coupling
\[
 \hat{H}_{\rm int}=-\chi \hat{a}_-^{\dagger}\hat{a}_- \sum_{j\in\mathcal{C}}\sum_{k=1}^2 k\ket{k}_j\bra{k}.
 \]
Then the conjugated evolution for a time $\tau=2\pi/3\chi$ from the initial state is 
\[
\hat{W}^{\dagger}e^{-i\hat{H}_{\rm int}\tau}\hat{W}\ket{\psi}\ket{\phi}_{\rm field}=\frac{1}{\sqrt{2}}(\ket{\psi}\ket{1}_+\hat{V}(\mathcal{C})\ket{\psi}\ket{1}_-)
\]
The measured polarization of the field mode in the bases $\ket{\pm_x}=(\ket{1}_+\pm\ket{1}_-)/\sqrt{2}$ and $\ket{\pm_y}=(\ket{1}_+\pm i\ket{1}_-)/\sqrt{2}$ is 
\[
\begin{array}{lll}
\langle \hat{\sigma}_{\rm field}^x\rangle=\Re[\langle \hat{V}(\mathcal{C})\rangle]\\
\langle \hat{\sigma}_{\rm field}^y\rangle=\Im[\langle \hat{V}(\mathcal{C})\rangle].
\end{array}
\]
A similar measurement of $V'(\mathcal{C})$ can be made with the same adaptations as in the geometric phase gate implementation.

While conceptually simpler, this method is in practice more challenging than the geometric phase gate because quantum control at the single photon level is needed, while in the former, only Gaussian states and operations on the bosonic mode are required.  

\subsection{Measurement of 't Hooft disorder operator}
The 't Hooft disorder operator acting on $n$ spins$-1$ particles for $U(1)$ QLM model is
\[
\hat{\Upsilon}(\varphi)=e^{-i\varphi \hat{S}_0^z}\otimes e^{i\varphi \hat{S}_1^z}\otimes \cdots \otimes e^{-i\varphi \hat{S}_{n-2}^z}\otimes e^{i\varphi \hat{S}_{n-1}^z}.
\]
where it is assumed the spins are ordered on a line in the dual lattice.
\subsubsection{Geometric phase gate}
The protocol is quite is quite similar to that for the Wilson loop measurement.  Choose the spin operator to be $\hat{O}=\sum_{j=0}^{n-1} (\ket{\hat{S}^z=1}_j\bra{\hat{S}^z=1}-\ket{\hat{S}^z=-1}_j\bra{\hat{S}^z=-1})$.  Now follow the same steps as in Eq. (\ref{controlledunitary}) but where $\theta\rightarrow -\theta$ for the even labelled spins, which can be obtained by picking $\chi\rightarrow -\chi$ in the interaction $H_{\rm int}$.

\begin{figure}[]
\centering
\includegraphics[width=0.75\columnwidth]{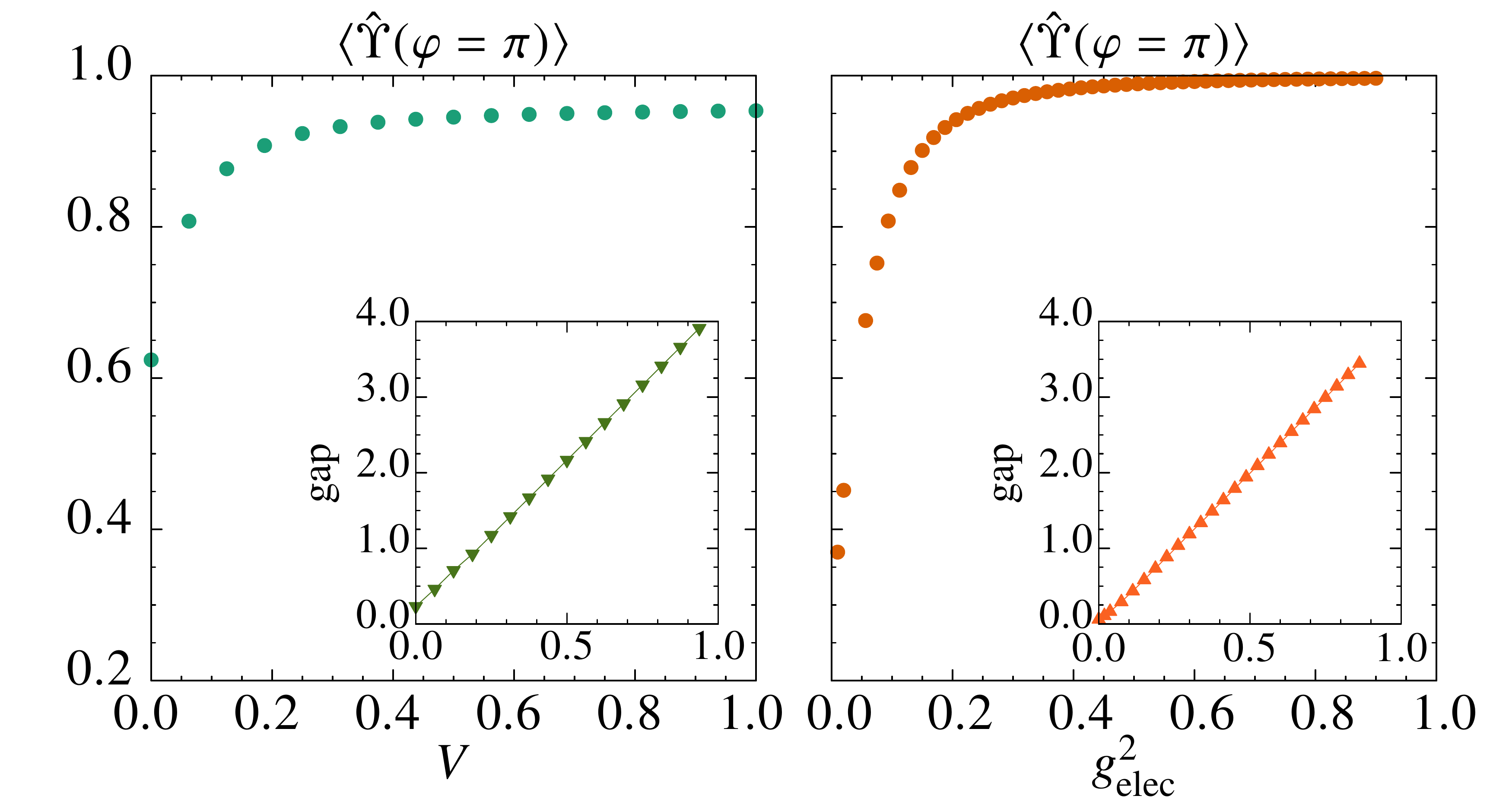}
\begin{caption}{\label{app1} Expectation value of the 't Hooft disorder parameter $\hat{\Upsilon}(\pi)$ and energy gap, in the gauge invariant subspace, for the ladder geometry in Fig. 1c of main text. (left panel) Value in the ground state of the two body Hamiltonian $\hat{H}_{\rm imp}$ with $J=1$ and $U=75$.  The energy gap is shown as a function of the on-site energy $V$. (right panel) Value in the ground state of the pure gauge model $\hat{H}_{\rm QLM}$. The energy gap is shown as a function of the electric coupling $g^2_{\rm elec}$, the magnetic coupling is fixed to the perturbative value $\frac{1}{g^2_{\rm mag}} \sim \frac{2J^{2}}{U} \to \frac{2}{75}$. Qualitatively this shows that the implementation Hamiltonian is well described by the effective pure gauge model for $J/U\ll 1$. Numerics were performed for a size $L=29$ rung ladder with $85$ spins total using DMRG with $300$ states and truncation error estimated at $<10^{-12}$. }
\end{caption}
\end{figure}

The measurement is easiest for the operator $\hat{\Upsilon}(\varphi=\pi)$. In that case, we can pick $\hat{O}=\sum_{j=0}^{n-1} \ket{\hat{S}^z=0}_j\bra{\hat{S}^z=0}$ and use the angle $\theta=\pi$ which means when needed one should evolve via the interaction $H_{\rm int}$ for a time $\tau=\pi/|\chi|$.   The unitary $U(\frac{\pi}{2},\pi,\frac{\pi}{2})=(-i)^{n}\hat{\Upsilon}(\pi)$, and performing the controlled unitary operator on the initial state
\[
\hat{M}(\frac{\pi}{2},\pi,\frac{\pi}{2})\ket{\psi}\ket{+_x}_A=\frac{1}{\sqrt{2}}(\ket{\psi}\ket{0}_A+(-i)^{n}\hat{\Upsilon}(\pi)\ket{\psi}\ket{1}_A).
 \]
The measured polarization of the ancilla along the $\hat{x}$ and $\hat{y}$ directions is 
\[
\begin{array}{lll}
\langle \hat{\sigma}_A^x\rangle(\frac{\pi}{2},\pi,\frac{\pi}{2})=\Re[\langle (-i)^{n}\hat{\Upsilon}(\pi)\rangle]\\
\langle \hat{\sigma}_A^y\rangle(\frac{\pi}{2},\pi,\frac{\pi}{2})=\Im[\langle (-i)^{n}\hat{\Upsilon}(\pi)\rangle],\\
\end{array}
\]
Hence we can perform a non-demolition measurement of the 't Hooft string $\hat{\Upsilon}(\pi)$.

\subsubsection{Single photon mediated gate}
\label{sp't Hooft}
The single photon mediated gate easily enables measurement of $\hat{\Upsilon}(\varphi)$ for any $\varphi$. Use the polarization dependent coupling which alternates sign on every other spin,
\[
 \hat{H}_{\rm int}=-\chi \hat{a}_-^{\dagger}\hat{a}_- \sum_{j=0}^{n-1}(-1)^j (\ket{\hat{S}^z=1}_j\bra{\hat{S}^z=1}-\ket{\hat{S}^z=-1}_j\bra{\hat{S}^z=-1}).
 \]
Then, following in the same manner as in the previous section where we measure the Wilson loop operator with a single photon mediated gate, the evolution from the initial state for a time $\tau=\varphi/|\chi|$ is 
\[
e^{-i\hat{H}_{\rm int}\tau}\ket{\psi}\ket{\phi}_{\rm field}=\frac{1}{\sqrt{2}}(\ket{\psi}\ket{1}_++\hat{\Upsilon}(\varphi)\ket{\psi}\ket{1}_-)
\]
The measured polarization of the field mode is 
\[
\begin{array}{lll}
\langle \hat{\sigma}_{\rm field}^x\rangle=\Re[\langle \hat{\Upsilon}(\varphi)\rangle]\\
\langle \hat{\sigma}_{\rm field}^y\rangle=\Im[\langle \hat{\Upsilon}(\varphi)\rangle].
\end{array}
\]
allowing for computing the real and imaginary parts of the expectation value.

One may ask how well the 't Hooft string order characterizes the order when using an engineered QLM model. In Fig. \ref{app1} we show that on a ladder geometry, qualitatively, the value of the 't Hooft string in the ground state of the exact 4 body Hamiltonian $\hat{H}_{QLM}$ of the main text is the same as that in the ground state of the engineered 2 body Hamiltonian $\hat{H}_{\rm imp}$. Moreover, in both versions, the system is gapped allowing for efficient ground state preparation.  

\subsection{Fidelity of the measurement}
\subsubsection{Effects due to cavity decay and spin depolarization}

The process fidelity for many-spin gates generated via coupling to bosonic channels was calculated in Ref. \cite{Brennen:09}.  For the uncontrolled geometric phase gate $U(\phi,\theta,\Omega)$ acting on qubits, it was shown that when the only form of decoherence is field decay at a rate $\kappa$, the process fidelity is lower bounded by $F=1-\frac{\pi \Omega\kappa}{|\chi|}(e^{-3\theta\kappa/2|\chi|}+e^{-\theta\kappa/2|\chi|})(1+\frac{\pi\kappa}{2|\chi|})$, where $\chi$ is the spin-field dispersive coupling. Including also collective depolarization of the spins at a rate $\gamma$ over a time period $t$ for the total gate, giving rise to an error probability $p=1-e^{-\gamma t}$, as well as independent depolarization at a rate $\gamma_i$ acting on each of the $n$ spins involved, giving rise independent depolarization error rate $p_i=1-e^{-\gamma_i \tau}$, the process fidelity is $F_{\rm pro}>(1-np_i-p)F$.

We can obtain a fidelity bound $F^{(gp)}_{\rm pro}(\theta,\Omega)$ for the geometric phase realization of controlled unitary operator $\hat{M}(\phi,\theta,\Omega)$ by a simple adaptation of the above calculation.  Coherent state displacements can be made very quickly with respect to other time scales in the system and the number of interaction gates in $\hat{M}(\phi,\theta,\Omega)$ is $7$, i.e., $4$ more than those needed for $U(\phi,\theta,\Omega)$ (assuming the initial state of the bosonic mode is vacuum).  Depolarization of the ancilla spin acts like collective decoherence in the system and we can assume it occurs at the same rate as independent error (i.e., $p=p_i=1-e^{-\gamma t}$).  For dispersive coupling the total time for the gate $\hat{M}$ is $t\approx (4\pi+6 \theta)/|\chi|$, where we have assumed that the maximum dispersive coupling to any basis state of the spins is $\chi$ so that each controlled rotation $R(\theta \hat{O})$ actually takes time $2\theta/|\chi|$. The entire process fidelity for measuring the many body operator, including implementing the controlled unitary $\hat{M}(\phi,\theta,\Omega)$ using the geometric phase gate on $n$ spins (including $1$ ancilla spin), followed by measurement of the ancilla is then
 \begin{equation}
 \begin{array}{lll}
F^{(gp)}_{\rm pro}(\theta,\Omega)>\eta_{A}\Big(1-n(1-e^{-(4\pi+6 \theta)\gamma/|\chi|})\Big) \Big(1-\frac{\pi\Omega\kappa}{|\chi|} (e^{-3\theta\kappa/|\chi|}+e^{-\theta\kappa/|\chi|})\big(1+\frac{\pi\kappa}{2|\chi|}\big)\Big).
\label{gpfidelity}
\end{array}
\end{equation}
where $\eta_{A}\leq 1$ describes finite detection fidelity of the ancilla spin.

For the Wilson loop measurement $\Omega=\pi/\sqrt{3}, \theta=2\pi/3$, while for the 't Hooft string $\hat{\Upsilon}(\pi)$ measurement $\Omega=\pi/2,\theta=\pi/2$.  The reason why we take $\theta=\pi/2$ for the latter is that in Eq. (\ref{gpfidelity}) we assumed that twice the field-spin interaction time was needed to accumulate the phase on one of the basis states while, as described in the section where we measure the 't Hooft disorder string operator with a single photon mediated gate, for measuring $\hat{\Upsilon}(\pi)$ only one basis state need interact with the field mode.

For a realization of the controlled unitary by a single photon gate the fidelity is enhanced significantly using feed forward control \cite{Brennen:09}.  During the gate the cavity trapping the field mode can be monitored for leakage by detectors.  A case of null detection improves the performance of the gate relative to an unmonitored cavity, and if a detection is seen then the gate can be repeated. The total process fidelity for measuring the many body operator is
  \begin{equation}
F^{(sp)}_{\rm pro}(\theta)>\eta_{p}\Big(1-n(1-e^{-\gamma\bar{t}(\theta)})\Big).
\label{spfidelity}
\end{equation}
where $\eta_{p}\leq 1$ describes finite single photon detection fidelity, and where the mean gate time is 
\begin{equation}
\bar{t}(\theta)=\frac{(1+e^{2\theta \kappa/|\chi|})(2\theta)^2\kappa/|\chi|^2}{2\theta \kappa/|\chi|+e^{2\theta \kappa/|\chi|}-1}.
\end{equation}
Here, for the Wilson loop measurement $\theta=2\pi/3$, while for measurement of the 't Hooft string $\hat{\Upsilon}(\varphi)$, $\theta=\varphi/2$.

\subsubsection{Effects of inhomogeneity}
Here we consider the effects of inhomogeneous coupling of spins to the cavity. This can be modeled by modifying the dispersive interaction as
 \begin{equation}
\hat{H}_{\rm int}=- \hat{a}^{\dagger}\hat{a} \sum_{j\in\mathcal{C}} \chi_j\sum_{k=1}^2 k\ket{k}_j\bra{k},
 \end{equation}
where the dispersive coupling strength $\chi$ may vary in space. To simplify the analysis, let's consider a system of $|\mathcal{C}|=n$ qubits and a target many body gate $U_{\rm target}=e^{-i\Omega Z^{\otimes n}}$ (without loss of generality we pick $n\bmod 4=1$). According to Eq. \ref{unitary}, the general form for the unitary operator generated using the geometric phase gate is $U(\phi,\theta,\Omega)= \exp(-i\Omega\sin(\theta\hat{O}+\phi))$ and we get our target unitary by picking $\phi=0$, $\theta=\frac{\pi}{2}$, and $\hat{O}=\sum_j \hat{Z}_j$ where $\hat{Z}=\ket{0}\bra{0}-\ket{1}\bra{1}$. This  could be generated by a uniform dispersive coupling: $\hat{H}_{\rm int}=- \hat{a}^{\dagger}\hat{a} \sum_{j} \chi_j \hat{Z}_j$ with $\chi_j=\chi \forall j$. In the presence of inhomogeneities, the actual unitary will be $U= \exp(-i\Omega\sin(\sum_j(\frac{\pi}{2}(1+\epsilon_j)\hat{Z}_j))$ where $\epsilon_j=(\chi_j-\chi)/\chi$, or equivalently
\[
\begin{array}{lll}
U(0,\frac{\pi}{2},\Omega)&=&\exp(-i\frac{\Omega}{2i}((iZ)^{\otimes n}\prod_j (\cos(\frac{\pi \epsilon_j}{2})+i\sin\frac{\pi \epsilon_j}{2}\hat{Z}_j)-(-iZ)^{\otimes n}\prod_j (\cos\frac{\pi \epsilon_j}{2}-i\sin\frac{\pi \epsilon_j}{2}\hat{Z}_j).
 \end{array}
 \]
If the value of the inhomogeneity is known, say by gate tomography, then we can adjust $\Omega\rightarrow \Omega/\prod_j \cos\frac{\pi \epsilon_j}{2}=\tilde{\Omega}$ by simply adjusting the magnitude of displacement operators during the geometric phase gate. Now
\[
\begin{array}{lll}
U(0,\frac{\pi}{2},\tilde{\Omega})&=&\exp(-i\frac{\Omega}{2i}((iZ)^{\otimes n}\prod_j (1+i\tan\frac{\pi \epsilon_j}{2}\hat{Z}_j)-(-iZ)^{\otimes n}\prod_j (1-i\tan\frac{\pi \epsilon_j}{2}\hat{Z}_j)\\
&=&U_{\rm target}\times \exp(i\Omega\sum_{a\neq b}\tan\frac{\pi \epsilon_a}{2}\tan\frac{\pi \epsilon_b}{2}\prod_{r\neq a,b}\hat{Z}_r)\\
&&\times \exp(-i\Omega\sum_{a\neq b\neq c\neq d}\tan\frac{\pi \epsilon_a}{2}\tan\frac{\pi \epsilon_b}{2}\tan\frac{\pi \epsilon_c}{2}\tan\frac{\pi \epsilon_d}{2}\prod_{r\neq a,b,c,d}\hat{Z}_r)\times \cdots.
 \end{array}
 \]
 The error $\mathcal{E}$ in the gate implementation due to inhomogeneities can be quantified as follows
 \[
\mathcal{E}\equiv|| U^{-1}_{\rm target}U(0,\frac{\pi}{2},\tilde{\Omega})-{\bf 1}||_{2}=||{\bf 1}-[e^{i\Omega\sum_{a\neq b}\tan\frac{\pi \epsilon_a}{2}\tan\frac{\pi \epsilon_b}{2}\prod_{r\neq a,b}\hat{Z}_r}][e^{-i\Omega\sum_{a\neq b\neq c\neq d}\tan\frac{\pi \epsilon_a}{2}\tan\frac{\pi \epsilon_b}{2}\tan\frac{\pi \epsilon_c}{2}\tan\frac{\pi \epsilon_d}{2}\prod_{r\neq a,b,c,d}\hat{Z}_r}]\cdots||_{2},
\]
where the norm on an operator $A$ is $||A||_{2}=\sup_x \frac{||Ax||_2}{||x||_2}$. Define $\epsilon=\max_j \epsilon_a$, then
\[
\begin{array}{lll}
\mathcal{E} & \leq & ||-i \sum_{j=1}^{\lfloor n \rfloor/2}(-1)^j\tan^{2 j}\frac{\pi \epsilon}{2}\sum_{\{a_1,\ldots a_{2j}\}}\prod\hat{Z}_{a_k}||_2\\
&\leq&\sum_{j=1}^{\lfloor n \rfloor/2}\tan^{2 j}\frac{\pi \epsilon}{2}\sum_{\{a_1,\ldots a_{2j}\}} ||\prod\hat{Z}_{a_k}||_2\\
&=&\sum_{j=1}^{\lfloor n \rfloor/2} {n\choose 2j} \tan^{2 j}\frac{\pi \epsilon}{2}\\
&=&\frac{1}{2}((\tan\frac{\pi \epsilon}{2}-1)^n+(\tan\frac{\pi \epsilon}{2}+1)^n-2).
\end{array}
\]
where the sum over members of the set $\{a_1,\ldots a_{2j}\}$ is over the choices of $2j$ distinct spin locations. For $\epsilon\ll 1$ the error is, $|| U^{-1}_{\rm target}U(0,\frac{\pi}{2},\tilde{\Omega})-{\bf 1}||_{2}\approx n(n-1)(\frac{\pi}{2})^2\epsilon^2/2$.  For gates with different rotation angles $\theta$, the error scales the same with the replacement $(\frac{\pi}{2})^2\rightarrow \theta^2$. The many body gates used for measurement of loop and strings in the spin$-1$ realization of quantum link models can be obtained by compositions of many body gates on qubit subspaces as described above.

\section{On gauge invariance and \emph{``dressed''} quantum states}

\begin{figure}[!h]
\centering
\includegraphics[width=\textwidth]{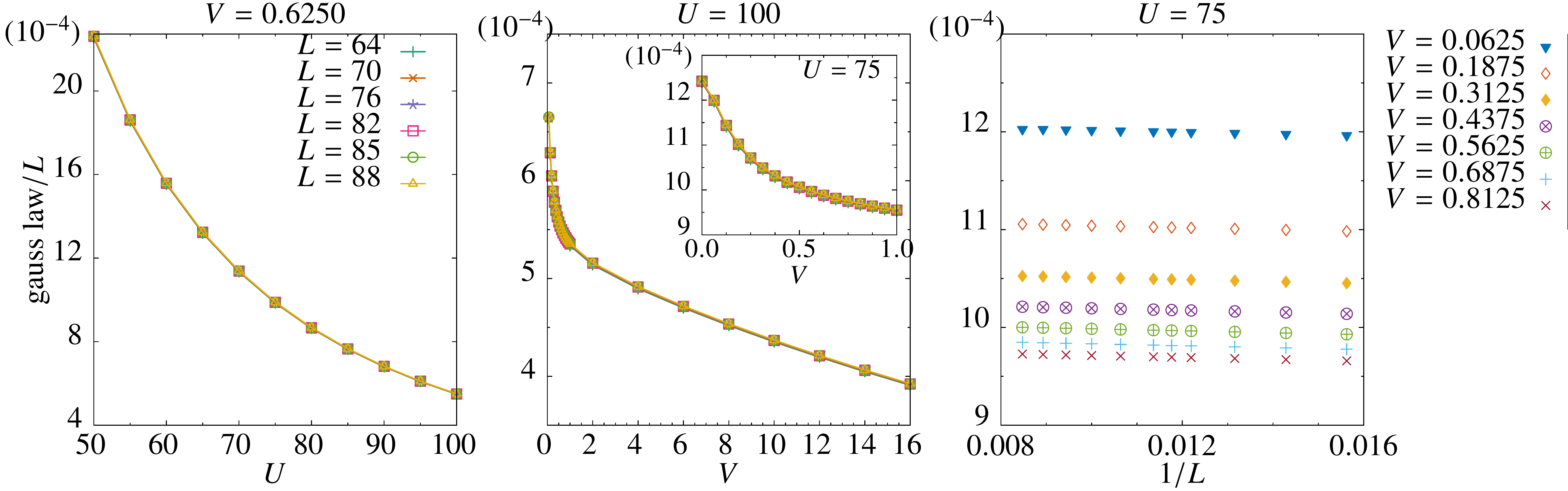}
\begin{caption}{\label{density} Gauss' law per lattice site in the ladder geometry shown in the main text. In the left panel, it is plot the quadratic dependence with the large scale penalty $U$. The middle panel shows the linear dependence of the Gauss' law with the local on-site energy $V$ at strong coupling, while in the inset, this dependence is slowly modified approaching the weak coupling limit. The right panel plots the independence of the Gauss' law density with the lattice size} 
\end{caption}
\end{figure}

In the main text, we show how an implementation based on superconducting circuits allows the emergence of a $U(1)$ gauge invariant QLM in the low energy sector. Because we are dealing with a system with local symmetries and local constants of motion, we need to perform a more careful analysis of the implementation by ``natural'' non-gauge invariant interactions and how this affects the Gauss' law in a real experiment.

The local Hilbert space of gauge invariant model is blocked diagonal in the different sectors characterized by the Gauss' law. In our case, the Gauss' law is nothing but the total magnetization of four spins ($S=1$) around every vertex in a lattice. The different gauge sectors are described by this value, i.e. $\hat{G}_\text{vertex} |m_\text{vertex} \rangle= \left( \sum_{i \in \text{vertex}} \hat{S}_{i}^{z} \right) |m_\text{vertex} \rangle =  m_\text{vertex} |m_\text{vertex} \rangle$. The ``charge-free'' or ``vacuum'' sector corresponds to the one with zero magnetization.

To simulate a gauge invariant model out of gauge variant interactions, we force the energy separation of the different gauge sectors with a large energy scale $U$. In perturbation theory, the gauge variant term will couple different sectors, but these processes are not energetically favorable, hence, in second order of perturbation, any dynamics that appear in the different gauge sectors are gauge invariant.

Following this approach, we break gauge invariance to introduce non-trivial interactions. To characterize this breaking, we plot in Fig.\ref{density}, the Gauss' law per lattice site as a function of the different parameters of the implemented Hamiltonian. The first panel shows that quadratic dependence of the Gauss' law with the large energy scale $U$. This fact is a direct consequence of second order perturbation theory. The panel with the on-site energy $V$ shows a linear dependence at least the middle and strong coupling limits, while there is a more complicated dependence in the weak coupling limit. Finally, we can see that Gauss' law per lattice site is independent of the lattice size in the last panel of Fig.\ref{density}.

Due to the fact of the small deviation of the Gauss' law from the ``charge-free'' or ``vacuum'' sector or the equivalence of the 't Hooft parameter between the implemented model and the gauge invariant model in the strongly coupled and intermediately coupled regimes, we can take a more active point of view to characterized perturbative processes. This second point of view is related with the \emph{``dressed''} quantum states that are used in quantum optics to describe interacting Hamiltonians. Within this second description, we follow the different quantum states as we turn on the perturbation $\frac{J}{U}$ and more concretely the different gauge sectors i.e., $|m_\text{vertex} \rangle \mapsto |\tilde{m}_\text{vertex} \rangle = |m_\text{vertex} \rangle +  \sum_{k}\alpha_{m,k}\left( \frac{J}{U} \right)  |k_\text{vertex} \rangle$. Due to the dressing or hybridization of the states, we will recover the second order Hamiltonian as an interacting one. In the same way, we can redefine the Gauss' operator: $\hat{G}_\text{vertex} \mapsto \hat{\tilde{G}}_\text{vertex} = \sum_{m} m_\text{vertex} |\tilde{m}_\text{vertex} \rangle  \langle \tilde{m}_\text{vertex}|$, so gauge invariance is recovered in this picture, at least, for the middle and strong coupling regimes. Obviously, a more detailed description is needed as we approach the weak coupling limit.

\end{widetext}


\begin{thebibliography}{99}
\bibitem{footnote} It was proven in Ref.\cite{Preskill}  that non-demolition measurements of space like non-Abelian Wilson loops are not physical because they would allow for faster than light signaling. However, Abelian Wilson loops such as considered here have no such obstruction provided the spins on the loop have access to a shared entanglement resource which here is the common cavity mode.
\end{thebibliography}
\end{document}